\documentclass[aps,prl,reprint,floatfix,groupedaddress,superscriptaddress,showpacs]{revtex4-1}

\usepackage{lipsum}
\usepackage{graphicx}
\usepackage{natbib}
\usepackage[utf8]{inputenc}
\usepackage{tikz}
\usepackage{amsmath}
\usepackage{empheq}
\usepackage{geometry}
\usepackage{amssymb}

\newcommand{\srni}{Sr$_3$NiIrO$_6$}

\begin{document}

\title{Giant gap in the magnon excitations of the quasi-1D chain compound \srni}

\author{W. Wu}
\affiliation{Department of Physics and Astronomy and London Centre for Nanotechnology, University College London, Gower Street, London WC1E 6BT, UK}

\author{D. T. Adroja}
\email{devashibhai.adroja@stfc.ac.uk}
\affiliation{ISIS Facility, STFC, Rutherford Appleton Laboratory, Chilton, Oxfordshire OX11 0QX, UK}
\affiliation{Highly Correlated Matter Research Group, Physics Department, University of Johannesburg, PO Box 524, Auckland Park 2006, South Africa}

\author{S. Toth}
\affiliation{Laboratory for Neutron Scattering, Paul Scherrer Institut (PSI), CH-5232 Villigen, Switzerland}
\affiliation{Laboratory for Quantum Magnetism, ICMP, Ecole Polytechnique Fédérale de Lausanne (EPFL), CH-1015 Lausanne, Switzerland}

\author{S. Rayaprol}
\affiliation{UGC-DAE CSR, Mumbai Center, R-5 Shed, BARC, Trombay, Mumbai 400085, India}

\author{E. V. Sampathkumaran}
\affiliation{Tata Institute of Fundamental Research, Homi Bhabha Road, Colaba, Mumbai 400005, India}

\date{\textrm{\today}}
\pacs{75.25.-j, 75.30.Cr, 75.30.Ds, 75.30.Gw, 75.40.Gb, 75.40.Mg, 75.47.Lx}

\begin{abstract}
Inelastic neutron scattering on the spin-chain compound \srni reveals gapped quasi-1D magnetic excitations. The observed one-magnon band between 29.5 and 39 meV consists of two dispersive modes. The spin wave spectrum could be well fitted with an antiferromagnetic anisotropic exchange model and single ion anisotropy on the Ni site. The extracted dominant anisotropic antiferromagnetic intra-chain exchange interaction between Ir and Ni ions are $J_z=19.5$ meV and $J_{xy}=12.1$ meV. These values justify previous electronic structure calculations, showing the importance of Ir spin orbit coupling on the electron correlations. The magnetic excitations survive up to 200 K well above the magnetic ordering temperature of $T_N \sim 70$ K, also indicating a quasi-1D nature of the magnetic interactions in \srni. Our results not only support the idea of the existence of a new temperature scale well above $T_N$, but also emphasize the need to consider new exchange paths complicated by the SOC, resulting in  additional characteristic temperatures in such spin-chain systems.
\end{abstract}

\maketitle

Low-dimensional systems, in which quantum mechanical wave function is confined in one or two spatial dimensions, exhibit some of the most interesting physical phenomena seen in condensed matter physics \cite{Bethe1931,Kubo1952}. One of these, (quasi-) one-dimensional (1D) spin chain is important for understanding low-dimensional magnetism \cite{Bethe1931,Kubo1952,Luther1975,Bose2007,Basu2013,Parkinson1985,Nightingale1986,ZurLoye2012,Nguyen1995,Vajenine1996,Stitzer2002,Basu2014} and useful for advancing quantum communications \cite{Bose2007,Basu2013}. The study of quantum spin chains has a long history, going back at least to the remarkable exact solution found by Bethe in 1931 for the spin-1/2 case \cite{Bethe1931}. Spin wave theory, developed by Anderson \cite{Anderson1959}, Kubo \cite{Kubo1952} and others in the 1950s \cite{Mattis1988,White1987,Oguchi1960,Dyson1956}, has given a clear physical picture of the behavior of three-dimensional (3D) antiferromagnets (AFM) \cite{Goldstone1962,Burgess2000}. However, 1D case remained rather mysterious as compared to the higher dimensions. Regarding 1D case, Haldane first suggested that integer-spin AFM Heisenberg spin chains have a finite gap, and only the half-odd-integer chains are gapless \cite{Haldane1983a,Haldane1983b}. 

\begin{figure}[!htb]
    \centering
	\includegraphics[width = \linewidth]{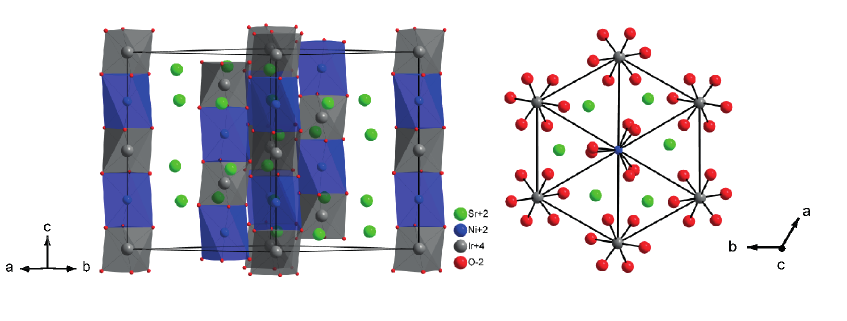}
	\caption{(Color online)  Crystal structure of \srni\ showing 1D Ni-Ir chains along c-axis. Right side, shows the hexagonal packing of chains viewed along the chain direction (i.e. in ab-plane). Dark gray,  blue, green and red balls show Ir, Ni, Sr and O atoms, respectively.}
	\label{fig:struct}
\end{figure}

Significant effort has been devoted over the past several decades to understand the behavior of frustrated quasi-1D spin systems \cite{Mutka1991}, which exhibit a rich variety of phases due to the enhanced quantum fluctuations in reduced dimensionality. The spin-chain systems with general formula A$_3$MM'O$_6$ (A denotes Sr, Ca, etc and M/M' denotes transition metals) have attracted much attention in recent years, due to their reduced dimensionality \cite{Basu2014,Sampathkumaran2004,Takubo2005,Flahaut2003,Hillier2011,Mikhailova2012,Sarkar2010,Ou2014,Yin2013,Jain2013,Agrestini2014,Agrestini2008,Wu2005,Chapon2009,Sampathkumaran2002,Sampathkumaran2004a,Sampathkumaran2004,Takubo2005,Sampathkumaran2007}. This structure consists of 1D chains that are oriented along the $c$-axis and arranged in a triangular lattice in the $ab$ plane (see Fig.\ \ref{fig:struct}). The chains are formed by alternating face-sharing MO$_6$ trigonal prism and M’O$_6$ octahedra, and intercalated by A$^{2+}$ cation, thus forming a hexagonal arrangement as shown in Fig.\ \ref{fig:struct}.

Among these spin chain systems, the Ca-based system such as Ca$_3$Co$_2$O$_6$ has been extensively investigated 
\cite{Jain2013,Agrestini2014,Agrestini2008,Sampathkumaran2002,Sampathkumaran2004,Sampathkumaran2004a,Takubo2005,Sampathkumaran2007}.  The present work is primarily motivated by the original report \cite{Paulose2008} that there exists a characteristic temperature, $T^*$, well above long-range magnetic ordering temperature due to incipient spin-chain order \cite{Bindu2009,Gohil2010}. Such investigations are however rare for Sr-based systems. The Sr-based spin chain system \cite{Hillier2011,Niazi2002,Rayaprol2004,Mohapatra2007} has become a hot topic in condensed matter physics recently owing to its chemical flexibility \cite{Nguyen1994}. A strong intra-chain exchange coupling, which may change sign depending on the metal atom, could exist in Sr$_3$MIrO$_6$ (M=metal) besides the strong spin-orbit coupling (SOC) of Ir (effective $S=1/2$) may lead to an anisotropic exchange interactions \cite{Sarkar2010,Ou2014} and spin anisotropy in the high-spin carrier such as Ni$^{2+}$ ($3d^8$, $S=1$).  

Sr$_3$MIrO$_6$, with M=Co, Cu, Ni, and Zn, have been studied previously \cite{Flahaut2003a,Hillier2011,Mikhailova2012a,Sarkar2010,Ou2014}. In this connection, magnetic investigations on some of these compounds can be found in our past literature  \cite{Niazi2002a,Rayaprol2004,Niazi2002a,Niazi2001}. In Ref.\ \cite{Yin2013a}, the resonant inelastic X-ray scattering (RIXS) data revealed gapped spin wave excitations in Sr$_3$CuIrO$_6$ that could be fitted using linear spin wave theory, where a spin-1/2 chain with an anisotropic exchange interactions induced by the SOC was adopted. When M=Zn (completely filled $3d$-shell), AFM spin-1/2 chains are expected for Sr$_3$ZnIrO$_6$ owing to the super-exchange mechanism between Ir atoms via oxygen ligands \cite{Nguyen1995a,Vajenine1996}. In contrast, for M=Ni ($S=1$ for Ni$^{2+}$, following the Hund’s rule), an alternating chain of spin-1/2 and spin-1 ions are formed along the $c$-axis \cite{Mikhailova2012a}. One would expect a ferromagnetic interaction between Ni and Ir spins owing to the orthogonality between $3d$ and $5d$ orbitals as was first shown by ab initio calculations \cite{Sarkar2010}. However, recent experimental and theoretical results have shown that the coupling is AFM \cite{Ou2014,Lefrancois2014a}. This might be due to the strong SOC, which may affect the exchange pathway, thus changing the sign of exchange interaction \cite{Ou2014}. A good understanding of the magnetic properties of \srni\ may require both high-resolution probing technique and spin wave calculations, in which the SOC is presented in the form of anisotropic exchange interaction and spin anisotropies. We have also carried out the magnetization and heat capacity measurements of \srni\ to characterize the sample quality \cite{[{See Supplementary Materials for technical details}]Supp}. 

In this Letter, we present a combination of INS measurements and spin-wave calculation for \srni (the first study of this type).  Polycrystalline sample of \srni\ was prepared by solid-state reactions of NiO, IrO$_2$ and SrCO$_3$ \cite{Supp}. The X-ray powder diffraction (XRD) study at 300 K shows that the \srni\ sample was single phase and crystallized in the space group R$\overline{3}$c (space group No.: 167). The INS measurements were performed between 5 and 300 K using the high count rate time-of-flight chopper spectrometer, MERLIN at the ISIS facility, UK. To reduce the neutron absorption problem from Ir, we filled the fine powder of \srni\  in a thin-Al envelop. The sample was cooled down to 5 K inside the He-exchange gas using a closed cycle refrigerator. The INS measurements were carried out with various incident neutrons energies: $E_i=15$, 80, 150 and 500 meV.

\begin{figure}[!htb]
    \centering
	\includegraphics[width = \linewidth]{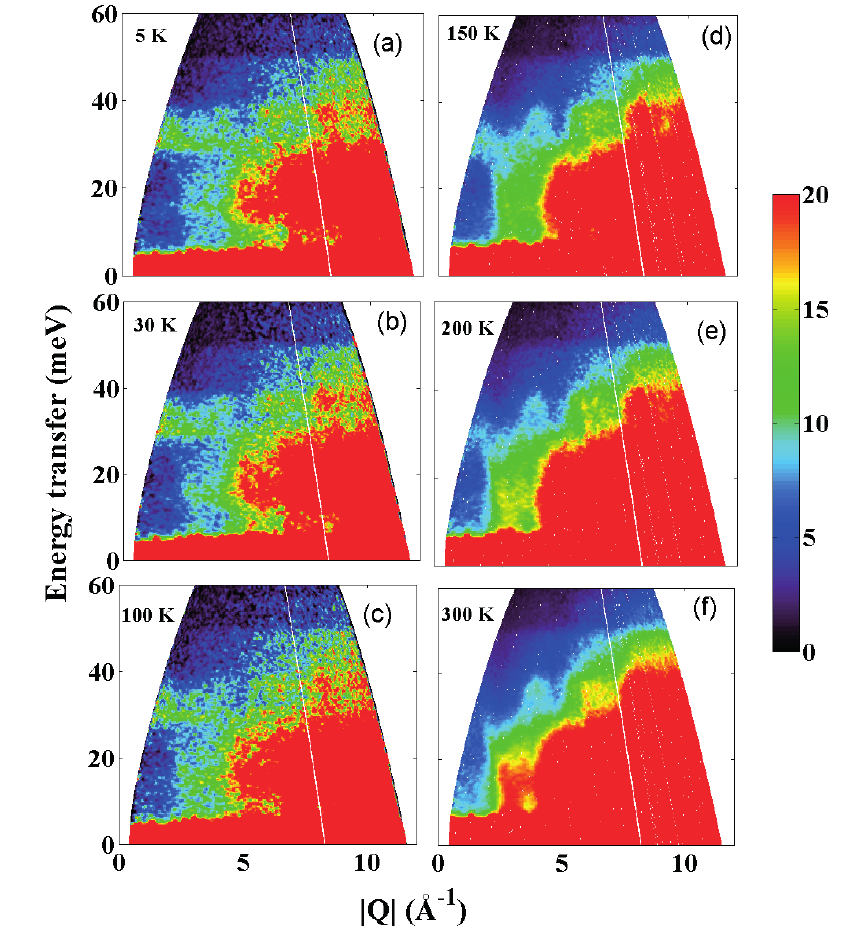}
	\caption{(Color online) Color coded inelastic neutron scattering intensity maps, energy transfer vs momentum transfer ($Q$) of \srni\ measured with an incident energy $E_i=80$ meV on MERLIN.}
	\label{fig:ins}
\end{figure}

The color-coded INS intensity maps of \srni\ measured at various temperatures between $T=5$ and 300 K with $E_i=80$ meV are shown in Fig.\ \ref{fig:ins}(a-f). At $T=5$ K, with the momentum transfer $|Q|$ below 4 \AA$^{-1}$, a strong scattering peak can be clearly observed between 29.5 meV and 39 meV (Fig.\ \ref{fig:ins}), without any sign of scattering below 30 meV at low-Q. Furthermore, we have not observed any clear magnetic signal above 50 meV in high-energy measurements up to 500 meV. The magnetic origin of the strong scattering in the 29.5-39 meV band is established by the initial reduction in the energy-integrated inelastic intensity as a function of $|Q|$ (see Fig.\ 2 in \cite{Supp}) up to 4 \AA$^{-1}$ caused by spin-wave excitation, and then an increase with $|Q|$ due to phonon scattering. The presence of the phonon scattering is revealed in the color maps (Fig.\ \ref{fig:ins}) by a large enhancement in the intensity between 10 meV and 40 meV at high-Q  ($|Q|\sim 8-10$ \AA$^{-1}$). We attribute the observed magnetic excitations between 29.5 and 39 meV for $|Q|$ below 4 \AA$^{-1}$ to spin wave scattering from Ni$^{2+}$ and Ir$^{4+}$ ions in the magnetically ordered state below 70 K \cite{Lefrancois2014a}, as expected. Amazingly, the spin wave excitations survive up to $T=200$ K (Fig.\ \ref{fig:ins}(a-e)). In addtion, the temperature hardly affects the magnon lifetime. This type of low-dimensional magnetic anomaly could be correlated to the characteristic temperature, $T^*$, proposed for this family of compounds \cite{Paulose2008}, which was subsequently confirmed by the observation of spin wave excitation in INS study of the single crystal and powder Ca$_3$Co$_2$O$_6$ \cite{Jain2013,Agrestini2014}. Sr$_3$NiRhO$_6$ (spin-1 and spin-1/2 alternating chain), similar to \srni, has also revealed a clear magnetic excitation near 20 meV up to 100 K, which is well above its $T_N=65$ K \cite{Rayaprol2004}. In addition, the INS study on Sr$_3$ZnIrO$_6$, which is formed by spin-1/2 chains, has showed spin wave excitations near 4 meV (5 meV) at 5 K, but the excitations disappear at $T_N=19$ K (16 K). The comparison between \srni\ and Sr$_3$ZnIrO$_6$ suggests that there exists a one-dimensional magnetic nature in the former whereas the latter lacks in it. 

This difference can be understood as follows. The presence of a giant spin gap and large zone-boundary energy in \srni, can be explained only if there exists strong anisotropic magnetic exchange interaction and/or single-ion anisotropy owing to SOC. This is further supported by the small spin gap observed in Sr$_3$ZnIrO$_6$ in which the SOC would have similar strength compared to that in \srni. In addition, the intra-chain and inter-chain Ir-Ir distances are very similar ($\sim 5.8$ \AA) in Sr$_3$ZnIrO$_6$, which is much larger than the nearest-neighboring Ni-Ir distance ($\sim2.7$ \AA) in \srni. Hence it is expected that the intra-chain interaction is comparable to the inter-chain interaction in Sr$_3$ZnIrO$_6$, while the intra-chain coupling is dominant in \srni. This could result in the disappearance of 1D magnetism in Sr$_3$ZnIrO$_6$.  

\begin{figure}[!htb]
    \centering
	\includegraphics[width = \linewidth]{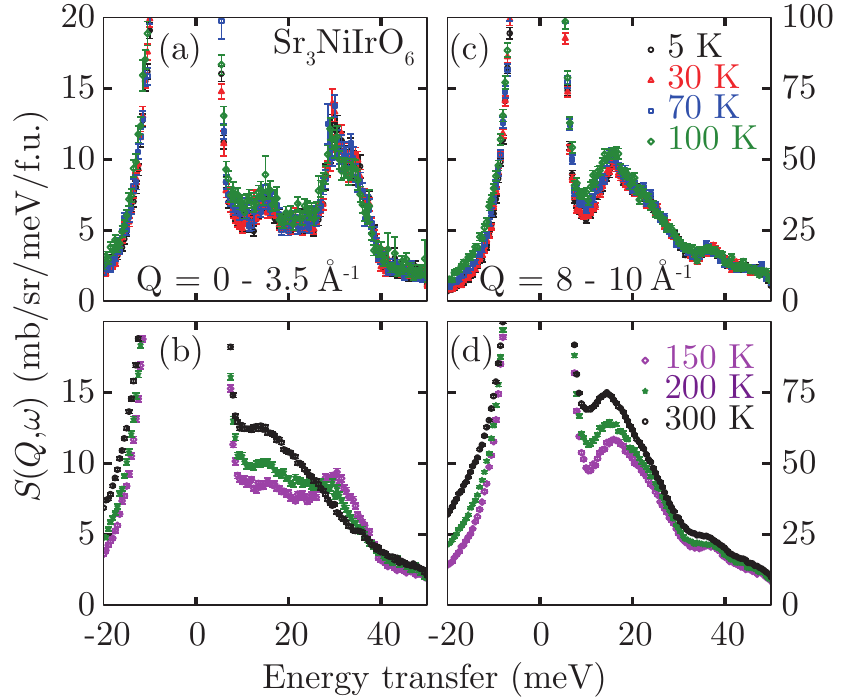}
	\caption{(Color online)  Q-integrated energy cuts at various temperatures from low-Q (0 to 3.5 \AA$^{-1}$) and high-Q (8 to 10 \AA$^{-1}$) for \srni.}
	\label{fig:cuts}
\end{figure}

\begin{figure}[!htb]
    \centering
	\includegraphics[width = \linewidth]{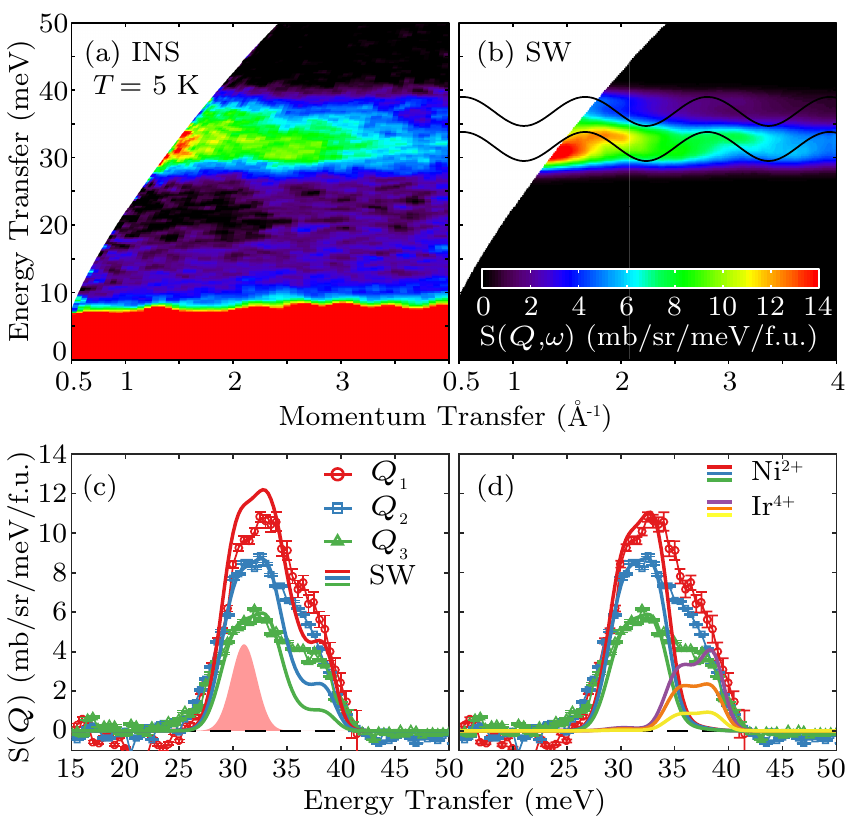}
	\caption{(Color online)  (a) The magnetic scattering at 5 K obtained after subtracting phonon scattering using 300 K data, strong scattering below 10 meV is due to the incoherent background. (b) The simulated spin wave scattering at 5 K using SpinW program [50] with exchange parameters, $J_z=19.5$ meV and $J_{xy}=12.14$ meV and a single ion anisotropy $D=-7.1$ meV. (c) 1D cuts of the magnetic scattering (symbols show experimental data) at 5 K for $Q_1=1.5-2$ \AA$^{-1}$ , $Q_2=2-3$ \AA$^{-1}$ and $Q_3=3-4$ \AA$^{-1}$ and simulated powder average spin wave (solid lines) using SpinW program with $J_z=19.5$ meV, $J_xy=12.14$ meV and $D=-7.1$ meV see (text). (d) shows the separate contributions (or decomposition) of the Ni and Ir calculated spin wave from (c). The solid Gaussian peak in (c) shows the instrumental energy resolution.}
	\label{fig:fit}
\end{figure}

To see the development of the inelastic magnetic scattering (at low-Q, 0 to 3.5 \AA$^{-1}$) and the phonon excitations (appear at high-Q, 8 to 10 \AA$^{-1}$), we have extracted Q-integrated energy cuts from the INS data as shown in Fig.\ \ref{fig:cuts}(a-d). At low-Q, between 5 and 100 K, we can see a strong inelastic peak near 30 meV with a shoulder near 35 meV (second peak) in Fig.\ \ref{fig:cuts}(a-b), which remains nearly temperature-independent up to 100 K. In high-Q cuts (Fig.\ \ref{fig:cuts}(c-d)) the absence of this 30-35 meV peak indicates that it is due to magnetic excitations and its intensity decreases with $Q$ as expected from the magnetic form factor of Ni$^{2+}$ and Ir$^{4+}$ ions (see Fig.\ 2(b) in \cite{Supp}). Furthermore, at high-Q, a strong peak near 15 meV (Fig.\ \ref{fig:cuts}(c-d)) is due to phonon excitations as its intensity is very weak at low-Q. Between 150 K and 300 K, at low-Q, the intensity at 30-35 meV peak decreases and almost disappears at 300 K, but that of 15 meV peak increases. To see magnetic excitations clearly, we have subtracted phonon scattering from the 5 K data using the data of 300 K. The magnetic scattering at 5 K thus derived in Fig. 4(a) reveals a clear magnetic excitation between 30 and 39 meV. To check the Q-dependent intensity of these excitations, we made Q-integrated cuts in three different ranges (Fig.\ \ref{fig:fit}(c)): $Q_1=1.5-2$ \AA$^{-1}$, $Q_2=2-3$ \AA$^{-1}$ and $Q_3=3-4$ \AA$^{-1}$. These cuts show a double-peak structure with a stronger peak around 33 meV and a weaker one around 37 meV.

The observed magnetic inelastic scattering can be well described using linear spin wave theory \cite{Bloch1930a,Slater1930}. The double-peak structure can be approximately attributed to spin waves on the Ir and Ni sites. The stronger lower-energy peak corresponds to spin waves on the Ni$^{2+}$ ions since the spin wave intensity is proportional to the spin quantum number. Besides, if the large spin gap is attributed to exchange anisotropy, the upper peak has to be Ir spin waves, due to the larger Weiss molecular field at the Ir sites created by the larger spin of the Ni$^{2+}$ ions. 

The published neutron diffraction data of \srni can be equally well fitted with two different solutions \cite{Lefrancois2014a} both with ordering wave vector $k=(0,0,1)$. In the first structure, the moments along the chains build up amplitude modulated AFM order, while in the second one only two third of the chains are ordered with AFM arrangement along the chains. In the partially ordered chain structure 2/3 of the moments would contribute to spin wave scattering and 1/3 to the diffuse scattering. In the spin wave calculations, we assume fully ordered chains of spin-1 and spin-1/2 ions and neglect inter-chain coupling. Owing to the SOC, we expect strongly anisotropic exchange interaction along the chain and a single-ion anisotropy for the spin-1 Ni site. Due to the threefold axis along the Ni-Ir chains (see Fig.1), the most general exchange matrix has two different diagonal elements, $J_z$ and $J_{xy}$ and the single-ion anisotropy can have only a $D_z$ component on the Ni-site. The observed magnetic structure suggests that $J_z$ is AFM and larger than $J_{xy}$. We neglect further neighbor interactions, based on similar 1D-chain models that have been explored for Sr$_3$CuIrO$_6$ \cite{Yin2013a} and Ca$_3$Co$_2$O$_6$ \cite{Jain2013}, both has given excellent agreements between the data and calculation and we could also achieve excellent result with the following simple model

\begin{align}
\label{eq:eq1}
\mathcal{H} =& \displaystyle\sum_{i} J_{xy}\left(S_i^xS_{i+1}^x+S_i^yS_{i+1}^y\right)  +J_zS_i^zS_{i+1}^z\\
&+ \sum_{i\in \textrm{Ni}} D_z S_i^zS_i^z\nonumber,
\end{align}

where $S_i$ with odd and even $i$ denotes Ir and Ni spins respectively. Using a two-sublattice single chain unit cell and effective Hamiltonian given by Eq.\ \ref{eq:eq1} the fitting of the spectrum is reduced to a single parameter fit, $D_z$ \cite{Supp}. To fit the observed spin wave spectrum, we have calculated the spin-spin correlation function and the neutron scattering cross section using the simulation package SpinW \cite{Toth2014}. Using $\omega_{min} = 29.5$ meV and $\omega_{max} = 39.0$ meV, the best fit gives a unique result:  $D_z=-7.15(5)$ meV, $J_z = 19.50(5)$ meV and $J_{xy} = 12.14(5)$ meV. The Q-integrated cuts convoluted with the instrumental energy resolution are shown on Fig.\ \ref{fig:fit}(c).  Fig.\ \ref{fig:fit}(d) shows the spin wave cross section localized on Ni and Ir sites separately. The calculated intensity includes the square of the g-factor of Ni$^{2+}$ ($S=1$, $L=3$, $J=4$, $g_{Ni}=5/4$) and Ir$^{4+}$ ($S=1/2$, $L=2$, $J=5/2$) ions and the magnetic form factor calculated from tabulated values and the g-factor \cite{[{}][{, eq. 7.28}]Lovesey1986}. The fit shows an excellent agreement with the data and also the Q-dependence is very well reproduced (see Fig.\ \ref{fig:fit}(b)). The only significant difference is the $Q$ dependent intensity of the upper Ir mode, where the measured intensity decreases slower as given by our magnetic form factor. This suggest that the Ir$^{4+}$ orbitals are significantly contracted in comparison to the free Ir$^{4+}$ ion. Ab-initio calculations would be necessary to precisely determine the magnetic form factor, however, this is beyond the scope of this paper. The calculated spin wave cross section agrees very well with the measured data in absolute units assuming fully ordered chains. Thus we can also exclude the partially ordered chain structure as a possible ground state, since it would reduce the spin wave cross section by 1/3. 

In conclusion, we have investigated \srni using inelastic neutron scattering, along with a spin-wave analysis. Our INS study reveals spin wave excitations with with a giant energy gap of 30 meV at 5 K. More strikingly, these gapped excitations survive up to a high temperature of 200 K, well above $T_N$, thus confirming the quasi-1D nature of the magnetic interaction. Our spin wave analysis has given an excellent description of the experimental data. The presence of giant spin gap, as compared to the very small spin gap in Sr$_3$ZnIrO$_6$ having only $5d$ magnetic ion (below 1.5 meV with ZB of 5 meV) reveals that mixed $3d$-$5d$ (or $3d$-$4d$) compounds can generate distinct exchange pathways and thus novel magnetic behaviors. Therefore, the present study can foster the research on the magnetic excitations in spin-chain systems to consider such hitherto unrealized factors, and would generate theoretical interest of the development of a more realistic model to understand the complex magnetic behavior of these systems.

\begin{acknowledgments}
We thank Prof. L.C. Chapon, Drs E. Lefrançois, P. McClarty, D. D. Khalyavin, A.D. Hillier, P. Manuel and W. Kockelmann for their involvement. We acknowledge interesting discussion with Profs. S. Lovesey and H. Wu. D.T.A. acknowledge financial assistance from CMPC-STFC grant number CMPC-09108. S. T. acknowledges funding from the European Community's Seventh Framework Programme (FP7/2007-2013) under grant agreement n.$^\circ$ 290605  (COFUND: PSI-FELLOW).
\end{acknowledgments}

\part*{\bf\Large\center SUPPLEMENTARY MATERIAL\\ Giant gap in the magnon excitations of the quasi-1D chain compound \srni}

\begin{center}

{W. Wu$^{1}$}
{D.T. Adroja$^{2,3}$} 
{S. Toth$^{4,5}$} 
{S. Rayaprol$^{6}$}
{E. V. Sampathkumaran$^{7}$}

\vspace{0.5cm}

$^1${\it Department of Physics and Astronomy and London Centre for Nanotechnology, University College London, Gower Street, London WC1E 6BT, UK} 

$^2${\it ISIS Facility, STFC, Rutherford Appleton Laboratory, Chilton, Oxfordshire OX11 0QX, UK}

$^3${\it Highly Correlated Matter Research Group, Physics Department, University of Johannesburg, \\PO Box 524, Auckland Park 2006, South Africa}

$^4${\it Laboratory for Neutron Scattering, Paul Scherrer Institut (PSI), CH-5232 Villigen, Switzerland}

$^5${\it Laboratory for Quantum Magnetism, ICMP, Ecole Polytechnique Fédérale de Lausanne (EPFL), CH-1015 Lausanne, Switzerland}

$^6${\it UGC-DAE CSR, Mumbai Center, R-5 Shed, BARC, Trombay, Mumbai 400085, India}

$^7${\it Tata Institute of Fundamental Research, Homi Bhabha Road, Colaba, Mumbai 400005, India}
\end{center}

\twocolumngrid

\section{Synthesis and sample characterization}

The polycrystalline sample of \srni\ ($\sim$6 g) was prepared by solid-state reactions in an Ar atmosphere at 1200 $^\circ$C. Stoichiometric powder mixtures of NiO  (Alfa Aesar, 99.999\% purity), IrO$_2$ (Umicore) and SrCO$_3$ (Alfa Aesar, 99.99\% purity) were annealed for 24 hours. The checking of the phase purity of the end product and the determination of unit cell parameters were carried out using an x-ray powder diffractometer (XRD) with Cu-K$\alpha$1 radiation at room temperature. Furthermore the quality of the sample was checked using neutron diffraction measurement at 100 K on the WISH diffractometer at ISIS facility of the Rutherford Appleton Laboratory (RAL) in the UK. The magnetization and heat capacity measurements were carried out using commercial MPMS and PPMS systems (Quantum Design) respectively in the temperature range of 2 and 300 K. 

\begin{ruledtabular}
\begin{table}[!htb]
	\centering
	\caption{Refined crystallographic parameters of \srni\ using the space group $R\overline{3}c$ at 2 K from neutron diffraction data \cite{Lefrancois2014asm}.}
	\label{tab:str}
	\begin{tabular}{ccccc}    
    \multicolumn{3}{l}{Unit cell dimensions:}&\multicolumn{2}{l}{$a=b=9.6089(4)$ \AA}\\
    \multicolumn{3}{l}{                     }&\multicolumn{2}{l}{$c=11.1678(4)$ \AA}\\
    \multicolumn{3}{l}{Angles:}&\multicolumn{2}{l}{$\alpha=\beta=90^\circ$,$\gamma=120^\circ$}\\
    \\
    Atom  &	Site&	$x$	     &	$y$	        &	$z$				\\        
    \hline
    Sr    & $18e$  & 0.3636(3) & 0 &  1/4 \\
    Ni    & $6a$   & 0         & 0 &  1/4 \\
    Ir    & $6b$   & 0         & 0 &    0 \\
 	O     & $36f$  & 0.1828(3) & 0.0247(4) & 0.1140(2) \\
	\end{tabular}
\end{table}
\end{ruledtabular}
%
%
\begin{ruledtabular}
\begin{table}[!htb]
	\centering
	\caption{Selected distances between atoms in \srni.}
	\label{tab:dist}
	\begin{tabular}{ccc}    
    Bond & \multicolumn{2}{c}{Bond length (\AA)} \\
         & Intra-chain    & Inter-chain           \\
    \hline
    Ir-Ir & 5.584 & 5.852 \\
    Ni-Ni & 5.584 & 5.852 \\
    Ni-Ir & 2.792 & 5.625 \\
    \hline
    Sr-O  & \multicolumn{2}{c}{2.477} \\
    Ir-O  & \multicolumn{2}{c}{2.014} \\
    Ni-O  & \multicolumn{2}{c}{2.183} \\
    O-O   & \multicolumn{2}{c}{2.710} \\
    Sr-Sr & \multicolumn{2}{c}{3.206} \\
    Sr-Ni & \multicolumn{2}{c}{3.494} \\
	\end{tabular}
\end{table}
\end{ruledtabular}

\begin{ruledtabular}
\begin{table}[!htb]
	\centering
	\caption{Bond angles on selected atoms in \srni.}
	\label{tab:bond}
	\begin{tabular}{cc}    
    Atoms & Angles ($^\circ$) \\
    \hline
    Ir-O-Ni & 80.59$^\circ$ \\
    Ir-O-Ir & 106.90$^\circ$ \\
    Ni-O-Ni & 110.80$^\circ$ \\
	\end{tabular}
\end{table}
\end{ruledtabular}
Our XRD analysis shows that the \srni\ sample was single phase and crystallized in the space group $R\overline{3}c$ (space group No.: 167) with $Z=6$ formula units per unit cell. The lattice parameters at $T=100$ K were determined from the neutron diffraction experiment \cite{Lefrancois2014sm}, values are shown in Table \ref{tab:str}, which are consistent with previous reports \cite{Nguyen1995sm,Vajenine1996sm,Flahaut2003sm,Mikhailova2012sm}. The different bond lengths and inter atomic distances are shown in Tab.\ \ref{tab:dist} and selected bond angles are given in Tab.\ \ref{tab:bond}. The crystal structure consists of chains aligned along the $c$-axis, formed by alternating face-sharing NiO$_6$ trigonal prisms and IrO$_6$ octahedra. The chains are arranged on a triangular lattice in the $ab$-plane. There are two Ir ions in the primitive unit at (0,0,0) and (0,0,1/2), and two Ni atom at (0,0,1/4) and (0,0,3/4). Taking into account the rhombohedral-lattice translations $t_1=(2/3,1/3,1/3)$, $t_2=(1/3,2/3,2/3)$, and identity, there are six Ir and six Ni sites in the conventional hexagonal unit-cell. The aforementioned symmetry leads to a shift of $1/6c$ between neighboring magnetic atoms along the chains. 

\section{Magnetization and Heat capacity}
 
\begin{figure}[!htb]
    \centering
	\includegraphics[width = \linewidth]{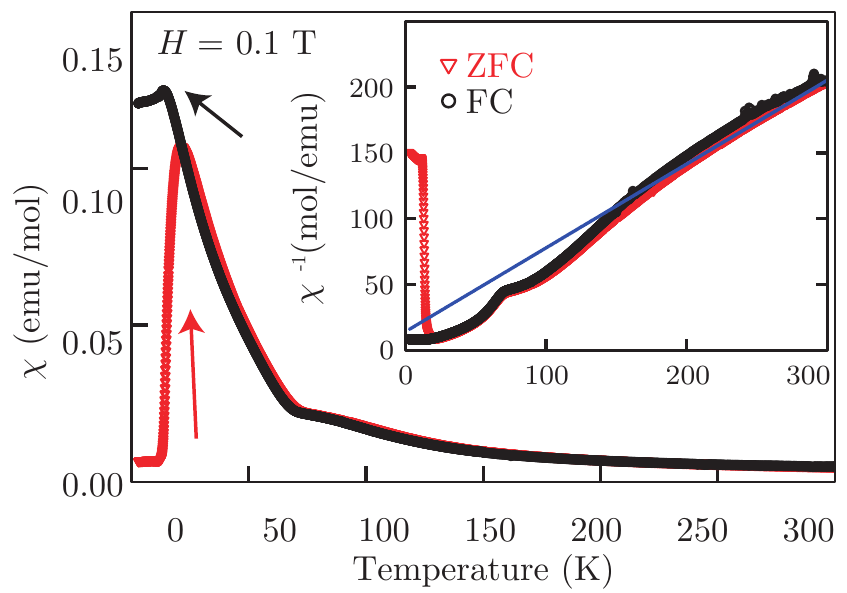}
	\caption{(Color online)  Magnetic susceptibility measured in zero-field cooled (red triangles) and field-cooled (black circles) conditions using 0.1 T field. The inset shows the inverse susceptibility versus temperature. The solid line denotes the linear Curie-Weiss fit between 150 and 300 K.}
	\label{fig:chi}
\end{figure}

Figure \ref{fig:chi} shows the magnetic susceptibility ($\chi$) of \srni as a function of temperature in zero-field-cooled (ZFC) and field-cooled (FC) conditions measured in an applied magnetic field of 0.1 T. Although both the ZFC and FC susceptibilities arise sharply below $T=70$ K, these exhibit considerably different behavior below $T=20$ K. While the ZFC signal sharply drops below 20 K, the FC susceptibility is almost constant below 20 K. Similar behavior was previously reported \cite{Flahaut2003asm,Mikhailova2012asm} and also observed in single crystals of \srni\ in external field parallel to the $c$-axis \cite{Lefrancois2014asm}. Furthermore, the spin-chain compound Sr$_3$NiRhO$_6$ ($T_N=65$ K) also exhibits very similar behavior in the ZFC and FC susceptibility \cite{Stitzer2002sm,Rayaprol2004sm,Mohapatra2007sm}. On the other hand, the susceptibilities of Sr$_3$ZnIrO$_6$ ($T_N=19$ K) \cite{Nguyen1995asm}\cite{Vajenine1996sm,Niazi2002sm,McClarty2014asm} and Sr$_3$ZnRhO$_6$ ($T_N=16$ K) \cite{Hillier2011sm,Rayaprol2004sm,Mohapatra2007sm} do not show any difference between ZFC and FC susceptibility below $T_N$. It is a general trend for the spin-chain systems of this family having two different magnetic atoms alternating along the chain to exhibit considerable different behavior in the temperature dependent susceptibility for ZFC and FC. The inverse susceptibility of \srni exhibits a Curie-Weiss (CW) behavior between 200 and 300 K with an effective total moment of $3.54(5)\mu_B$ per formula-unit and paramagnetic CW temperature of $-22.1(3)$ K for ZFC data (see inset of Fig.\ \ref{fig:chi}). However these values should be taken as rough estimates, due to the different temperature dependence of the paramagnetic susceptibility of Ir and Ni ions. The negative sign of the CW temperature indicates dominant antiferromagnetic interactions. Furthermore, $\chi$ exhibits a broad maximum below 100 K, indicating the low-dimensional nature (or due to frustration) of the magnetic interactions in \srni.
%
%
The measured ac-susceptibility has a peak in the real part ($\chi'$) at 20 K \cite{Lefrancois2014asm}, which shows strong frequency dependence (i.e. peak position increases with frequency) and the time dependent magnetization follows logarithmic decay at 20 K. These results show the importance of magnetic frustration due to the triangular nature of the crystal structure. 

\begin{figure}[!htb]
    \centering
	\includegraphics[width = \linewidth]{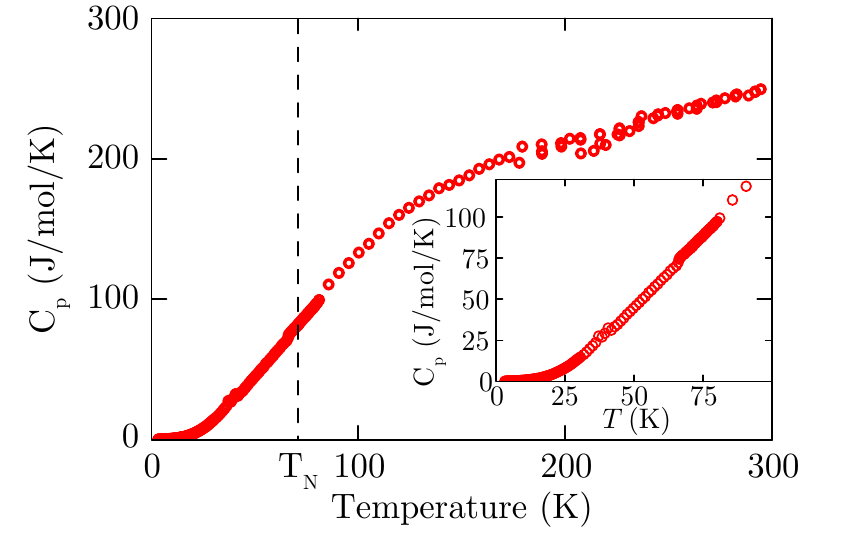}
	\caption{(Color online) Heat capacity versus temperature of \srni\ . The inset shows the heat capacity vs temperature plot in an expanded scale.}
	\label{fig:heat}
\end{figure}

We have also measured the heat capacity of \srni\ between 2 K and 300 K (see Figure \ref{fig:heat}) and haven’t found any clear signature of magnetic phase transition down to 2 K, even though the neutron diffraction study clearly reveals the present of magnetic Bragg peaks below 70 K \cite{Lefrancois2014asm}, confirming AFM magnetic order below 70 K. On the other hand, the neutron diffraction data did not show any clear change in the magnetic structure between 2 K and 70 K. It should be noted that the heat capacity of Sr$_3$NiRhO$_6$ \cite{Rayaprol2004sm,Adroja2014sm}, did not reveal any magnetic phase transition either, but those of Sr$_3$ZnIrO$_6$ \cite{Niazi2002sm,McClarty2014asm} and Sr$_3$ZnRhO$_6$ \cite{Hillier2011sm} exhibit a clear $\lambda$-type anomaly at the magnetic ordering temperature. This might indicate that the absence of the $\lambda$-type anomaly in \srni\ is strongly related to the low-dimensional nature in magnetic structure. 
%
%
The observed value of the heat capacity for \srni\ at 300 K is 254 J$\cdot$mol$^{-1}$K$^{-1}$, which is in a good agreement with that expected from the lattice contribution based on Dulong and Petit law \cite{Ashcroft1976sm} (for n-atom molecules $Cp\sim 3nR=274.4$ J$\cdot$mol$^{-1}$K$^{-1}$), showing that magnetic correlations are completely lost at room temperature.

\begin{figure}[!htb]
    \centering
	\includegraphics[width = \linewidth]{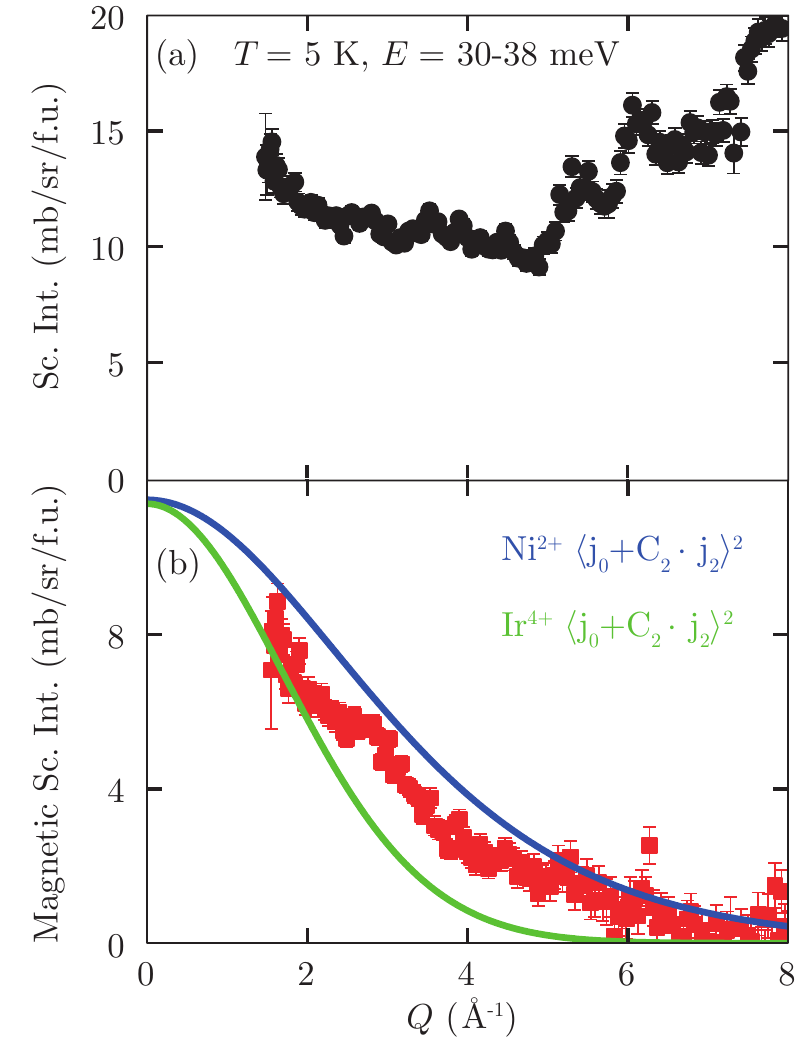}
	\caption{(a) Scattering intensity as a function of momentum transfer of \srni\ integrated in energy between 30 and 38meV measured at 5 K. (b) Magnetic scattering intensity only, obtained by subtracting phonon background. The solid lines show the magnetic form factor for Ni$^{2+}$ (blue line) and Ir$^{4+}$ (green line) scaled to the data.}
	\label{fig:ff}
\end{figure}

\begin{figure}[!htb]
    \centering
	\includegraphics[width = \linewidth]{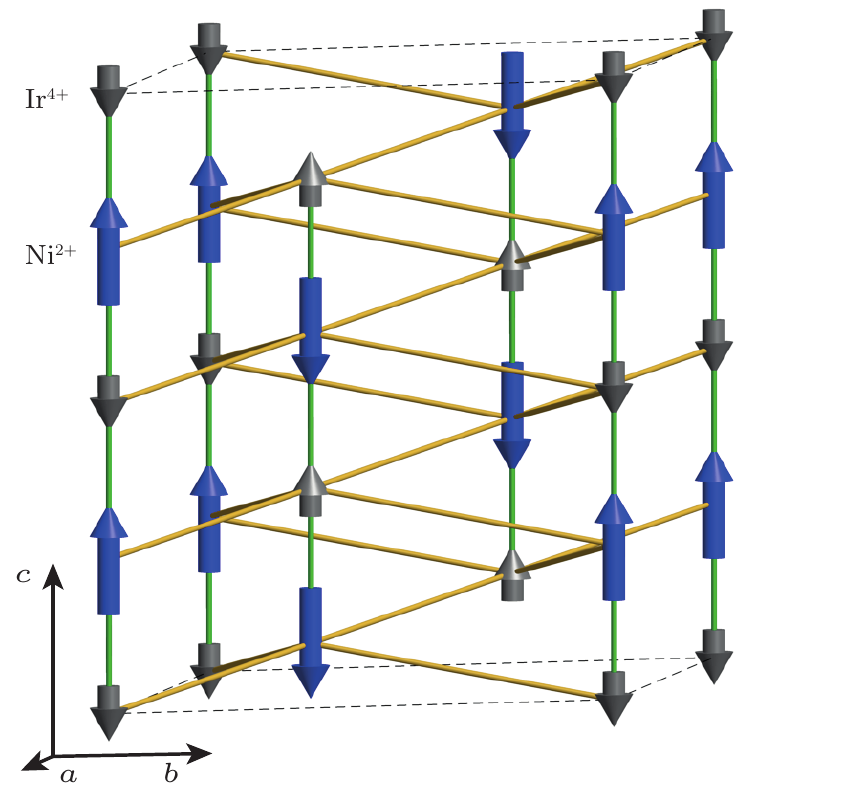}
	\caption{One of the possible magnetic structure of \srni\ stabilized at low temperature in a commensurate phase with propagation vector $k=(0,0,1)$. Green and orange lines denote intrachain and interchain couplings respectively. Structure of symmetry $P\overline{3}c'1$  corresponding to global phase  $\varphi=0$. For global phase $\varphi=\pi/6$ there is a possibility to have zero moment on Ni and Ir for the Ni-Ir chain at $(x,y,z)$ \cite{Lefrancois2014asm}.}
	\label{fig:Js}
\end{figure}

\section{Spin Wave calculations}

We propose a simple magnetic Hamiltonian, where we take into account only interactions between nearest neighbor magnetic atoms along the $c$-axis and single ion anisotropy of the Ni$^{2+}$ ions. This model gives a perfect fit to the inelastic neutron scattering (INS) powder data, this also justifies that the interchain interactions can be neglected. Due to the rhombohedral-translations every Ni-Ir chain are equivalent to a single chain with two atoms. We denote the Ir and Ni atoms with sublattice $a$ and $b$ respectively. We restrict our spin Hamiltonian using the crystallographic space group. The point group symmetry $D_3$ of the Ni atom allows only a single anisotropy parameter along the c-axis denoted by $D_z$. The allowed matrix elements of the Ni-Ir bond is determined by the point group symmetry of the center of the bond that is $C_3$. This allows anisotropic exchange, with different $J_z$ and $J_{xy}$ values, also Dzyaloshinskii-Moriya interaction is allowed parallel to the $c$-axis that we neglect here. With the above assumptions we propose the following effective spin Hamiltonian:
\begin{align}
\mathcal{H}  &=\displaystyle\sum_i J_z \hat{s}_{a,i}^z \hat{s}_{b,i}^z + J_{xy}\left(\hat{s}_{a,i}^x \hat{s}_{b,i}^x + \hat{s}_{a,i}^y \hat{s}_{b,i}^y \right)+ \nonumber \\
             +&\sum_i J_z \hat{s}_{a,i}^z \hat{s}_{b,i+1}^z + J_{xy}\left( \hat{s}_{a,i}^x \hat{s}_{b,i+1}^x + \hat{s}_{a,i}^y \hat{s}_{b,i+1}^y \right)+ \nonumber \\
             +&\sum_i D_z \left(\hat{s}_{b,i}^z\right)^2.
             \label{eq1}
\end{align}

The first two sums describe the anisotropic exchange interaction on both the Ni and Ir sites, while the last term describes the single-ion anisotropy of the Ni sites. In the observed magnetic structure the spins point along the $z$-axis AF order along the chain. The classical ground state of Eq.\ \ref{eq1} is identical to this if $J_z$ is positive and $|J_z|>|J_{xy}|$ for zero anisotropy. Assuming the AF ground state, we calculated the spin wave dispersion using the Holstein-Primakoff transformation \cite{Holstein1940sm} to reduce the Hamiltonian to non-interacting bosons (magnons) and applied the Bogoliubov transformation to diagonalize the quadratic form \cite{Bogoliubov1947sm}. Since there are two sublattices, we obtained two $\omega_{1,2}(Q)$ spin wave modes dispersing along the chain direction:
\begin{widetext}
\begin{align}
\omega_{1,2}(Q) = \left| \pm \sqrt{D_z^2s_b^2+2D_zJ_zs_b(s_a+s_b)+Jz^2(s_a+s_b)^2-2J_{xy}^2s_as_b(1+\cos(Q\cdot c/2))} + D_zs_b+J_z(s_a-s_b) \right|.
\end{align}
\end{widetext}
where $Q$ is the momentum of the magnon in \AA$^{-1}$ units, $c$ is the lattice parameter. In order to fit the measured spectrum, we calculated the bottom and top energies of the spin wave band ($\omega_{b}$ and $\omega_{t}$). If  $D_z<\omega_{t}/4$ we got:
\begin{align}
J_z =& \omega_{t}/2 \\
J_{xy}^2 =& D_z(\omega_{t}+\omega_{b})+1/4\omega_{t}^2-1/2\omega_{b}^2-1/4\omega_{b}\omega_{t}, \nonumber
\end{align}
while if $D_z>\omega_{t}/4$ we got:
\begin{align}
J_z   =& \omega_{t}-2D_z \\
J_{xy}^2 =& -2D_z(\omega_{t}+\omega_{b})+\omega_{t}^2-1/2\omega_{b}^2+1/2\omega_{b}\omega_{t}. \nonumber
\end{align}

If we determine the bottom and the top of the spin wave band from the INS data, there is only a single free parameter $D_z$ to be fitted. By calculating the powder spin wave spectrum by averaging out sufficiently enough random sample orientations for a series of $D_z$ values, we could determine $D_z$ by a direct comparison of the simulation and data.

An alternative fitting of the data is possible if we assume that we only observe the lower spin wave band. Thus there is an upper branch that we missed either due to its small INS cross section within our measured momentum transfer and energy range or it was above our maximum measured incident neutron energy ($E_i=500$ meV). Detection of spin waves at high energies is much harder due to the cutoff of the INS intensity at higher $Q$ values because of the magnetic form factor. We simulated this scenario within our spin wave model. The highest possible energy of the upper band is 75 meV assuming negative $D_z$, this belongs to $D_z=0$ meV, $J_z=37.5$ meV and $J_{xy}=20$ meV. In this case the gap is induced only by the difference between $J_z$ and $J_{xy}$. However in this case the lower band in the Q-integrated cuts would be symmetric as a function of energy, which is not consistent with the data, where the inelastic peak shows a shoulder higher energies. In addition, the Q-dependence of the lower magnon band would show the Ni form factor mainly, which doesn’t fit the data either, where the upper shoulder has clearly different Q-dependence than the peak intensity at 35 meV, see Fig.\ 4(c-d) in the manuscript. Thus we can exclude that there is an upper undetected spin wave branch with great confidence.


\begin{thebibliography}{59}%
\makeatletter
\providecommand \@ifxundefined [1]{%
 \@ifx{#1\undefined}
}%
\providecommand \@ifnum [1]{%
 \ifnum #1\expandafter \@firstoftwo
 \else \expandafter \@secondoftwo
 \fi
}%
\providecommand \@ifx [1]{%
 \ifx #1\expandafter \@firstoftwo
 \else \expandafter \@secondoftwo
 \fi
}%
\providecommand \natexlab [1]{#1}%
\providecommand \enquote  [1]{``#1''}%
\providecommand \bibnamefont  [1]{#1}%
\providecommand \bibfnamefont [1]{#1}%
\providecommand \citenamefont [1]{#1}%
\providecommand \href@noop [0]{\@secondoftwo}%
\providecommand \href [0]{\begingroup \@sanitize@url \@href}%
\providecommand \@href[1]{\@@startlink{#1}\@@href}%
\providecommand \@@href[1]{\endgroup#1\@@endlink}%
\providecommand \@sanitize@url [0]{\catcode `\\12\catcode `\$12\catcode
  `\&12\catcode `\#12\catcode `\^12\catcode `\_12\catcode `\%12\relax}%
\providecommand \@@startlink[1]{}%
\providecommand \@@endlink[0]{}%
\providecommand \url  [0]{\begingroup\@sanitize@url \@url }%
\providecommand \@url [1]{\endgroup\@href {#1}{\urlprefix }}%
\providecommand \urlprefix  [0]{URL }%
\providecommand \Eprint [0]{\href }%
\providecommand \doibase [0]{http://dx.doi.org/}%
\providecommand \selectlanguage [0]{\@gobble}%
\providecommand \bibinfo  [0]{\@secondoftwo}%
\providecommand \bibfield  [0]{\@secondoftwo}%
\providecommand \translation [1]{[#1]}%
\providecommand \BibitemOpen [0]{}%
\providecommand \bibitemStop [0]{}%
\providecommand \bibitemNoStop [0]{.\EOS\space}%
\providecommand \EOS [0]{\spacefactor3000\relax}%
\providecommand \BibitemShut  [1]{\csname bibitem#1\endcsname}%
\let\auto@bib@innerbib\@empty
\bibitem [{\citenamefont {Bethe}(1931)}]{Bethe1931}%
  \BibitemOpen
  \bibfield  {author} {\bibinfo {author} {\bibfnamefont {H.}~\bibnamefont
  {Bethe}},\ }\href {\doibase 10.1007/BF01341708} {\bibfield  {journal}
  {\bibinfo  {journal} {Z. Phys.}\ }\textbf {\bibinfo {volume}
  {71}},\ \bibinfo {pages} {205} (\bibinfo {year} {1931})}\BibitemShut
  {NoStop}%
\bibitem [{\citenamefont {Kubo}(1952)}]{Kubo1952}%
  \BibitemOpen
  \bibfield  {author} {\bibinfo {author} {\bibfnamefont {R.}~\bibnamefont
  {Kubo}},\ }\href {\doibase 10.1103/PhysRev.87.568} {\bibfield  {journal}
  {\bibinfo  {journal} {Phys. Rev.}\ }\textbf {\bibinfo {volume} {87}},\
  \bibinfo {pages} {568} (\bibinfo {year} {1952})}\BibitemShut {NoStop}%
\bibitem [{\citenamefont {Luther}\ and\ \citenamefont
  {Peschel}(1975)}]{Luther1975}%
  \BibitemOpen
  \bibfield  {author} {\bibinfo {author} {\bibfnamefont {A.}~\bibnamefont
  {Luther}}\ and\ \bibinfo {author} {\bibfnamefont {I.}~\bibnamefont
  {Peschel}},\ }\href {\doibase 10.1103/PhysRevB.12.3908} {\bibfield  {journal}
  {\bibinfo  {journal} {Phys. Rev. B}\ }\textbf {\bibinfo {volume} {12}},\
  \bibinfo {pages} {3908} (\bibinfo {year} {1975})}\BibitemShut {NoStop}%
\bibitem [{\citenamefont {Bose}(2007)}]{Bose2007}%
  \BibitemOpen
  \bibfield  {author} {\bibinfo {author} {\bibfnamefont {S.}~\bibnamefont
  {Bose}},\ }\href {\doibase 10.1080/00107510701342313} {\bibfield  {journal}
  {\bibinfo  {journal} {Contemp. Phys.}\ }\textbf {\bibinfo {volume} {48}},\
  \bibinfo {pages} {13} (\bibinfo {year} {2007})}\BibitemShut {NoStop}%
\bibitem [{\citenamefont {Basu}\ \emph {et~al.}(2013)\citenamefont {Basu},
  \citenamefont {Iyer}, \citenamefont {Singh},\ and\ \citenamefont
  {Sampathkumaran}}]{Basu2013}%
  \BibitemOpen
  \bibfield  {author} {\bibinfo {author} {\bibfnamefont {T.}~\bibnamefont
  {Basu}}, \bibinfo {author} {\bibfnamefont {K.~K.}\ \bibnamefont {Iyer}},
  \bibinfo {author} {\bibfnamefont {K.}~\bibnamefont {Singh}}, \ and\ \bibinfo
  {author} {\bibfnamefont {E.~V.}\ \bibnamefont {Sampathkumaran}},\ }\href
  {\doibase 10.1038/srep03104} {\bibfield  {journal} {\bibinfo  {journal} {Sci.
  Rep.}\ }\textbf {\bibinfo {volume} {3}},\ \bibinfo {pages} {3104} (\bibinfo
  {year} {2013})}\BibitemShut {NoStop}%
\bibitem [{\citenamefont {Parkinson}\ and\ \citenamefont
  {Bonner}(1985)}]{Parkinson1985}%
  \BibitemOpen
  \bibfield  {author} {\bibinfo {author} {\bibfnamefont {J.~B.}\ \bibnamefont
  {Parkinson}}\ and\ \bibinfo {author} {\bibfnamefont {J.~C.}\ \bibnamefont
  {Bonner}},\ }\href {\doibase 10.1103/PhysRevB.32.4703} {\bibfield  {journal}
  {\bibinfo  {journal} {Phys. Rev. B}\ }\textbf {\bibinfo {volume} {32}},\
  \bibinfo {pages} {4703} (\bibinfo {year} {1985})}\BibitemShut {NoStop}%
\bibitem [{\citenamefont {Nightingale}\ and\ \citenamefont
  {Bl\"{o}te}(1986)}]{Nightingale1986}%
  \BibitemOpen
  \bibfield  {author} {\bibinfo {author} {\bibfnamefont {M.~P.}\ \bibnamefont
  {Nightingale}}\ and\ \bibinfo {author} {\bibfnamefont {H.~W.~J.}\
  \bibnamefont {Bl\"{o}te}},\ }\href {\doibase 10.1103/PhysRevB.33.659}
  {\bibfield  {journal} {\bibinfo  {journal} {Phys. Rev. B}\ }\textbf {\bibinfo
  {volume} {33}},\ \bibinfo {pages} {659} (\bibinfo {year} {1986})}\BibitemShut
  {NoStop}%
\bibitem [{\citenamefont {zur Loye}\ \emph {et~al.}(2012)\citenamefont {zur
  Loye}, \citenamefont {Zhao}, \citenamefont {Bugaris},\ and\ \citenamefont
  {Chance}}]{ZurLoye2012}%
  \BibitemOpen
  \bibfield  {author} {\bibinfo {author} {\bibfnamefont {H.-C.}\ \bibnamefont
  {zur Loye}}, \bibinfo {author} {\bibfnamefont {Q.}~\bibnamefont {Zhao}},
  \bibinfo {author} {\bibfnamefont {D.~E.}\ \bibnamefont {Bugaris}}, \ and\
  \bibinfo {author} {\bibfnamefont {W.~M.}\ \bibnamefont {Chance}},\ }\href
  {\doibase 10.1039/c1ce05788j} {\bibfield  {journal} {\bibinfo  {journal}
  {CrystEngComm}\ }\textbf {\bibinfo {volume} {14}},\ \bibinfo {pages} {23}
  (\bibinfo {year} {2012})}\BibitemShut {NoStop}%
\bibitem [{\citenamefont {Nguyen}\ and\ \citenamefont {zur
  Loye}(1995{\natexlab{a}})}]{Nguyen1995}%
  \BibitemOpen
  \bibfield  {author} {\bibinfo {author} {\bibfnamefont {T.}~\bibnamefont
  {Nguyen}}\ and\ \bibinfo {author} {\bibfnamefont {H.-C.}\ \bibnamefont {zur
  Loye}},\ }\href {\doibase 10.1006/jssc.1995.1277} {\bibfield  {journal}
  {\bibinfo  {journal} {J. Solid State Chem.}\ }\textbf {\bibinfo {volume}
  {117}},\ \bibinfo {pages} {300} (\bibinfo {year}
  {1995}{\natexlab{a}})}\BibitemShut {NoStop}%
\bibitem [{\citenamefont {Vajenine}\ \emph {et~al.}(1996)\citenamefont
  {Vajenine}, \citenamefont {Hoffmann},\ and\ \citenamefont {{Zur
  Loye}}}]{Vajenine1996}%
  \BibitemOpen
  \bibfield  {author} {\bibinfo {author} {\bibfnamefont {G.~V.}\ \bibnamefont
  {Vajenine}}, \bibinfo {author} {\bibfnamefont {R.}~\bibnamefont {Hoffmann}},
  \ and\ \bibinfo {author} {\bibfnamefont {H.~C.}\ \bibnamefont {{Zur Loye}}},\
  }\href {\doibase 10.1016/0301-0104(95)00335-5} {\bibfield  {journal}
  {\bibinfo  {journal} {Chem. Phys.}\ }\textbf {\bibinfo {volume} {204}},\
  \bibinfo {pages} {469} (\bibinfo {year} {1996})}\BibitemShut {NoStop}%
\bibitem [{\citenamefont {Stitzer}\ \emph {et~al.}(2002)\citenamefont
  {Stitzer}, \citenamefont {Henley}, \citenamefont {Claridge}, \citenamefont
  {zur Loye},\ and\ \citenamefont {Layland}}]{Stitzer2002}%
  \BibitemOpen
  \bibfield  {author} {\bibinfo {author} {\bibfnamefont {K.}~\bibnamefont
  {Stitzer}}, \bibinfo {author} {\bibfnamefont {W.}~\bibnamefont {Henley}},
  \bibinfo {author} {\bibfnamefont {J.}~\bibnamefont {Claridge}}, \bibinfo
  {author} {\bibfnamefont {H.-C.}\ \bibnamefont {zur Loye}}, \ and\ \bibinfo
  {author} {\bibfnamefont {R.}~\bibnamefont {Layland}},\ }\href {\doibase
  10.1006/jssc.2001.9463} {\bibfield  {journal} {\bibinfo  {journal} {J. Solid
  State Chem.}\ }\textbf {\bibinfo {volume} {164}},\ \bibinfo {pages} {220}
  (\bibinfo {year} {2002})}\BibitemShut {NoStop}%
\bibitem [{\citenamefont {Basu}\ \emph {et~al.}(2014)\citenamefont {Basu},
  \citenamefont {Iyer}, \citenamefont {Singh}, \citenamefont {Mukherjee},
  \citenamefont {Paulose},\ and\ \citenamefont {Sampathkumaran}}]{Basu2014}%
  \BibitemOpen
  \bibfield  {author} {\bibinfo {author} {\bibfnamefont {T.}~\bibnamefont
  {Basu}}, \bibinfo {author} {\bibfnamefont {K.}~\bibnamefont {Iyer}}, \bibinfo
  {author} {\bibfnamefont {K.}~\bibnamefont {Singh}}, \bibinfo {author}
  {\bibfnamefont {K.}~\bibnamefont {Mukherjee}}, \bibinfo {author}
  {\bibfnamefont {P.}~\bibnamefont {Paulose}}, \ and\ \bibinfo {author}
  {\bibfnamefont {E.}~\bibnamefont {Sampathkumaran}},\ }\href {\doibase
  10.1063/1.4895699} {\bibfield  {journal} {\bibinfo  {journal} {Appl. Phys.
  Lett.}\ }\textbf {\bibinfo {volume} {105}},\ \bibinfo {pages} {102912}
  (\bibinfo {year} {2014})}\BibitemShut {NoStop}%
\bibitem [{\citenamefont {Anderson}(1959)}]{Anderson1959}%
  \BibitemOpen
  \bibfield  {author} {\bibinfo {author} {\bibfnamefont {P.~W.}\ \bibnamefont
  {Anderson}},\ }\href {\doibase 10.1103/PhysRev.115.2} {\bibfield  {journal}
  {\bibinfo  {journal} {Phys. Rev.}\ }\textbf {\bibinfo {volume} {115}},\
  \bibinfo {pages} {2} (\bibinfo {year} {1959})}\BibitemShut {NoStop}%
\bibitem [{\citenamefont {Mattis}(1988)}]{Mattis1988}%
  \BibitemOpen
  \bibfield  {author} {\bibinfo {author} {\bibfnamefont {D.~C.}\ \bibnamefont
  {Mattis}},\ }\href@noop {} {\emph {\bibinfo {title} {{Theory of Magnetism
  I}}}}\ (\bibinfo  {publisher} {Springer Verlag},\ \bibinfo {year}
  {1988})\BibitemShut {NoStop}%
\bibitem [{\citenamefont {White}(1987)}]{White1987}%
  \BibitemOpen
  \bibfield  {author} {\bibinfo {author} {\bibfnamefont {R.~M.}\ \bibnamefont
  {White}},\ }\href@noop {} {\emph {\bibinfo {title} {{Quantum theory of
  Magnetism}}}}\ (\bibinfo  {publisher} {Springer Verlag},\ \bibinfo {year}
  {1987})\BibitemShut {NoStop}%
\bibitem [{\citenamefont {Oguchi}(1960)}]{Oguchi1960}%
  \BibitemOpen
  \bibfield  {author} {\bibinfo {author} {\bibfnamefont {T.}~\bibnamefont
  {Oguchi}},\ }\href {\doibase 10.1103/PhysRev.117.117} {\bibfield  {journal}
  {\bibinfo  {journal} {Phys. Rev.}\ }\textbf {\bibinfo {volume} {117}},\
  \bibinfo {pages} {117} (\bibinfo {year} {1960})}\BibitemShut {NoStop}%
\bibitem [{\citenamefont {Dyson}(1956)}]{Dyson1956}%
  \BibitemOpen
  \bibfield  {author} {\bibinfo {author} {\bibfnamefont {F.~J.}\ \bibnamefont
  {Dyson}},\ }\href {\doibase 10.1103/PhysRev.102.1217} {\bibfield  {journal}
  {\bibinfo  {journal} {Phys. Rev.}\ }\textbf {\bibinfo {volume} {102}},\
  \bibinfo {pages} {1217} (\bibinfo {year} {1956})}\BibitemShut {NoStop}%
\bibitem [{\citenamefont {Goldstone}\ \emph {et~al.}(1962)\citenamefont
  {Goldstone}, \citenamefont {Salam},\ and\ \citenamefont
  {Weinberg}}]{Goldstone1962}%
  \BibitemOpen
  \bibfield  {author} {\bibinfo {author} {\bibfnamefont {J.}~\bibnamefont
  {Goldstone}}, \bibinfo {author} {\bibfnamefont {A.}~\bibnamefont {Salam}}, \
  and\ \bibinfo {author} {\bibfnamefont {S.}~\bibnamefont {Weinberg}},\ }\href
  {\doibase 10.1103/PhysRev.127.965} {\bibfield  {journal} {\bibinfo  {journal}
  {Phys. Rev.}\ }\textbf {\bibinfo {volume} {127}},\ \bibinfo {pages} {965}
  (\bibinfo {year} {1962})}\BibitemShut {NoStop}%
\bibitem [{\citenamefont {Burgess}(2000)}]{Burgess2000}%
  \BibitemOpen
  \bibfield  {author} {\bibinfo {author} {\bibfnamefont {C.~P.}\ \bibnamefont
  {Burgess}},\ }\href {\doibase 10.1016/S0370-1573(99)00111-8} {\bibfield
  {journal} {\bibinfo  {journal} {Phys. Reports}\ }\textbf {\bibinfo {volume}
  {330}},\ \bibinfo {pages} {193} (\bibinfo {year} {2000})}\BibitemShut
  {NoStop}%
\bibitem [{\citenamefont {Haldane}(1983{\natexlab{a}})}]{Haldane1983a}%
  \BibitemOpen
  \bibfield  {author} {\bibinfo {author} {\bibfnamefont {F.}~\bibnamefont
  {Haldane}},\ }\href {\doibase 10.1016/0375-9601(83)90631-X} {\bibfield
  {journal} {\bibinfo  {journal} {Phys. Lett. A}\ }\textbf {\bibinfo {volume}
  {93}},\ \bibinfo {pages} {464} (\bibinfo {year}
  {1983}{\natexlab{a}})}\BibitemShut {NoStop}%
\bibitem [{\citenamefont {Haldane}(1983{\natexlab{b}})}]{Haldane1983b}%
  \BibitemOpen
  \bibfield  {author} {\bibinfo {author} {\bibfnamefont {F.~D.~M.}\
  \bibnamefont {Haldane}},\ }\href {\doibase 10.1103/PhysRevLett.50.1153}
  {\bibfield  {journal} {\bibinfo  {journal} {Phys. Rev. Lett.}\ }\textbf
  {\bibinfo {volume} {50}},\ \bibinfo {pages} {1153} (\bibinfo {year}
  {1983}{\natexlab{b}})}\BibitemShut {NoStop}%
\bibitem [{\citenamefont {Mutka}\ \emph {et~al.}(1991)\citenamefont {Mutka},
  \citenamefont {Payen}, \citenamefont {Molini\'{e}}, \citenamefont
  {Soubeyroux}, \citenamefont {Colombet},\ and\ \citenamefont
  {Taylor}}]{Mutka1991}%
  \BibitemOpen
  \bibfield  {author} {\bibinfo {author} {\bibfnamefont {H.}~\bibnamefont
  {Mutka}}, \bibinfo {author} {\bibfnamefont {C.}~\bibnamefont {Payen}},
  \bibinfo {author} {\bibfnamefont {P.}~\bibnamefont {Molini\'{e}}}, \bibinfo
  {author} {\bibfnamefont {J.}~\bibnamefont {Soubeyroux}}, \bibinfo {author}
  {\bibfnamefont {P.}~\bibnamefont {Colombet}}, \ and\ \bibinfo {author}
  {\bibfnamefont {A.}~\bibnamefont {Taylor}},\ }\href {\doibase
  10.1103/PhysRevLett.67.497} {\bibfield  {journal} {\bibinfo  {journal} {Phys.
  Rev. Lett.}\ }\textbf {\bibinfo {volume} {67}},\ \bibinfo {pages} {497}
  (\bibinfo {year} {1991})}\BibitemShut {NoStop}%
\bibitem [{\citenamefont {Sampathkumaran}\ \emph
  {et~al.}(2004{\natexlab{a}})\citenamefont {Sampathkumaran}, \citenamefont
  {Fujiwara}, \citenamefont {Rayaprol}, \citenamefont {Madhu},\ and\
  \citenamefont {Uwatoko}}]{Sampathkumaran2004}%
  \BibitemOpen
  \bibfield  {author} {\bibinfo {author} {\bibfnamefont {E.~V.}\ \bibnamefont
  {Sampathkumaran}}, \bibinfo {author} {\bibfnamefont {N.}~\bibnamefont
  {Fujiwara}}, \bibinfo {author} {\bibfnamefont {S.}~\bibnamefont {Rayaprol}},
  \bibinfo {author} {\bibfnamefont {P.~K.}\ \bibnamefont {Madhu}}, \ and\
  \bibinfo {author} {\bibfnamefont {Y.}~\bibnamefont {Uwatoko}},\ }\href
  {\doibase 10.1103/PhysRevB.70.014437} {\bibfield  {journal} {\bibinfo
  {journal} {Phys. Rev. B}\ }\textbf {\bibinfo
  {volume} {70}},\ \bibinfo {pages} {014437} (\bibinfo {year}
  {2004}{\natexlab{a}})}\BibitemShut {NoStop}%
\bibitem [{\citenamefont {Takubo}\ \emph {et~al.}(2005)\citenamefont {Takubo},
  \citenamefont {Mizokawa}, \citenamefont {Hirata}, \citenamefont {Son},
  \citenamefont {Fujimori}, \citenamefont {Topwal}, \citenamefont {Sarma},
  \citenamefont {Rayaprol},\ and\ \citenamefont {Sampathkumaran}}]{Takubo2005}%
  \BibitemOpen
  \bibfield  {author} {\bibinfo {author} {\bibfnamefont {K.}~\bibnamefont
  {Takubo}}, \bibinfo {author} {\bibfnamefont {T.}~\bibnamefont {Mizokawa}},
  \bibinfo {author} {\bibfnamefont {S.}~\bibnamefont {Hirata}}, \bibinfo
  {author} {\bibfnamefont {J.~Y.}\ \bibnamefont {Son}}, \bibinfo {author}
  {\bibfnamefont {A.}~\bibnamefont {Fujimori}}, \bibinfo {author}
  {\bibfnamefont {D.}~\bibnamefont {Topwal}}, \bibinfo {author} {\bibfnamefont
  {D.~D.}\ \bibnamefont {Sarma}}, \bibinfo {author} {\bibfnamefont
  {S.}~\bibnamefont {Rayaprol}}, \ and\ \bibinfo {author} {\bibfnamefont
  {E.~V.}\ \bibnamefont {Sampathkumaran}},\ }\href {\doibase
  10.1103/PhysRevB.71.073406} {\bibfield  {journal} {\bibinfo  {journal} {Phys.
  Rev. B}\ }\textbf {\bibinfo {volume} {71}},\
  \bibinfo {pages} {073406} (\bibinfo {year} {2005})}\BibitemShut {NoStop}%
\bibitem [{\citenamefont {Flahaut}\ \emph
  {et~al.}(2003{\natexlab{a}})\citenamefont {Flahaut}, \citenamefont {Hebert},
  \citenamefont {Maignan}, \citenamefont {Hardy}, \citenamefont {Martin},
  \citenamefont {Hervieu}, \citenamefont {Costes}, \citenamefont {Raquet},\
  and\ \citenamefont {Broto}}]{Flahaut2003}%
  \BibitemOpen
  \bibfield  {author} {\bibinfo {author} {\bibfnamefont {D.}~\bibnamefont
  {Flahaut}}, \bibinfo {author} {\bibfnamefont {S.}~\bibnamefont {Hebert}},
  \bibinfo {author} {\bibfnamefont {A.}~\bibnamefont {Maignan}}, \bibinfo
  {author} {\bibfnamefont {V.}~\bibnamefont {Hardy}}, \bibinfo {author}
  {\bibfnamefont {C.}~\bibnamefont {Martin}}, \bibinfo {author} {\bibfnamefont
  {M.}~\bibnamefont {Hervieu}}, \bibinfo {author} {\bibfnamefont
  {M.}~\bibnamefont {Costes}}, \bibinfo {author} {\bibfnamefont
  {B.}~\bibnamefont {Raquet}}, \ and\ \bibinfo {author} {\bibfnamefont {J.~M.}\
  \bibnamefont {Broto}},\ }\href {\doibase 10.1140/epjb/e2003-00283-3}
  {\bibfield  {journal} {\bibinfo  {journal} {Eur. Phys. J. B
  }\ }\textbf {\bibinfo {volume} {35}},\ \bibinfo {pages} {317} (\bibinfo
  {year} {2003}{\natexlab{a}})}\BibitemShut {NoStop}%
\bibitem [{\citenamefont {Hillier}\ \emph {et~al.}(2011)\citenamefont
  {Hillier}, \citenamefont {Adroja}, \citenamefont {Kockelmann}, \citenamefont
  {Chapon}, \citenamefont {Rayaprol}, \citenamefont {Manuel}, \citenamefont
  {Michor},\ and\ \citenamefont {Sampathkumaran}}]{Hillier2011}%
  \BibitemOpen
  \bibfield  {author} {\bibinfo {author} {\bibfnamefont {A.~D.}\ \bibnamefont
  {Hillier}}, \bibinfo {author} {\bibfnamefont {D.~T.}\ \bibnamefont {Adroja}},
  \bibinfo {author} {\bibfnamefont {W.}~\bibnamefont {Kockelmann}}, \bibinfo
  {author} {\bibfnamefont {L.~C.}\ \bibnamefont {Chapon}}, \bibinfo {author}
  {\bibfnamefont {S.}~\bibnamefont {Rayaprol}}, \bibinfo {author}
  {\bibfnamefont {P.}~\bibnamefont {Manuel}}, \bibinfo {author} {\bibfnamefont
  {H.}~\bibnamefont {Michor}}, \ and\ \bibinfo {author} {\bibfnamefont {E.~V.}\
  \bibnamefont {Sampathkumaran}},\ }\href {\doibase 10.1103/PhysRevB.83.024414}
  {\bibfield  {journal} {\bibinfo  {journal} {Phys. Rev. B
  }\ }\textbf {\bibinfo {volume} {83}},\ \bibinfo {pages} {024414}
  (\bibinfo {year} {2011})}\BibitemShut {NoStop}%
\bibitem [{\citenamefont {Mikhailova}\ \emph
  {et~al.}(2012{\natexlab{a}})\citenamefont {Mikhailova}, \citenamefont
  {Schwarz}, \citenamefont {Senyshyn}, \citenamefont {Bell}, \citenamefont
  {Skourski}, \citenamefont {Ehrenberg}, \citenamefont {Tsirlin}, \citenamefont
  {Agrestini}, \citenamefont {Rotter}, \citenamefont {Reichel}, \citenamefont
  {Chen}, \citenamefont {Hu}, \citenamefont {Li}, \citenamefont {Li},\ and\
  \citenamefont {Tjeng}}]{Mikhailova2012}%
  \BibitemOpen
  \bibfield  {author} {\bibinfo {author} {\bibfnamefont {D.}~\bibnamefont
  {Mikhailova}}, \bibinfo {author} {\bibfnamefont {B.}~\bibnamefont {Schwarz}},
  \bibinfo {author} {\bibfnamefont {A.}~\bibnamefont {Senyshyn}}, \bibinfo
  {author} {\bibfnamefont {a.~M.~T.}\ \bibnamefont {Bell}}, \bibinfo {author}
  {\bibfnamefont {Y.}~\bibnamefont {Skourski}}, \bibinfo {author}
  {\bibfnamefont {H.}~\bibnamefont {Ehrenberg}}, \bibinfo {author}
  {\bibfnamefont {a.~a.}\ \bibnamefont {Tsirlin}}, \bibinfo {author}
  {\bibfnamefont {S.}~\bibnamefont {Agrestini}}, \bibinfo {author}
  {\bibfnamefont {M.}~\bibnamefont {Rotter}}, \bibinfo {author} {\bibfnamefont
  {P.}~\bibnamefont {Reichel}}, \bibinfo {author} {\bibfnamefont {J.~M.}\
  \bibnamefont {Chen}}, \bibinfo {author} {\bibfnamefont {Z.}~\bibnamefont
  {Hu}}, \bibinfo {author} {\bibfnamefont {Z.~M.}\ \bibnamefont {Li}}, \bibinfo
  {author} {\bibfnamefont {Z.~F.}\ \bibnamefont {Li}}, \ and\ \bibinfo {author}
  {\bibfnamefont {L.~H.}\ \bibnamefont {Tjeng}},\ }\href {\doibase
  10.1103/PhysRevB.86.134409} {\bibfield  {journal} {\bibinfo  {journal} {Phys.
  Rev. B}\ }\textbf {\bibinfo {volume} {86}},\ \bibinfo {pages} {134409}
  (\bibinfo {year} {2012}{\natexlab{a}})}\BibitemShut {NoStop}%
\bibitem [{\citenamefont {Sarkar}\ \emph {et~al.}(2010)\citenamefont {Sarkar},
  \citenamefont {Kanungo},\ and\ \citenamefont {Saha-Dasgupta}}]{Sarkar2010}%
  \BibitemOpen
  \bibfield  {author} {\bibinfo {author} {\bibfnamefont {S.}~\bibnamefont
  {Sarkar}}, \bibinfo {author} {\bibfnamefont {S.}~\bibnamefont {Kanungo}}, \
  and\ \bibinfo {author} {\bibfnamefont {T.}~\bibnamefont {Saha-Dasgupta}},\
  }\href {\doibase 10.1103/PhysRevB.82.235122} {\bibfield  {journal} {\bibinfo
  {journal} {Phys. Rev. B}\ }\textbf {\bibinfo {volume} {82}},\ \bibinfo
  {pages} {235122} (\bibinfo {year} {2010})}\BibitemShut {NoStop}%
\bibitem [{\citenamefont {Ou}\ and\ \citenamefont {Wu}(2014)}]{Ou2014}%
  \BibitemOpen
  \bibfield  {author} {\bibinfo {author} {\bibfnamefont {X.}~\bibnamefont
  {Ou}}\ and\ \bibinfo {author} {\bibfnamefont {H.}~\bibnamefont {Wu}},\ }\href
  {\doibase 10.1038/srep04609} {\bibfield  {journal} {\bibinfo  {journal} {Sci.
  Rep.}\ }\textbf {\bibinfo {volume} {4}},\ \bibinfo {pages} {4609} (\bibinfo
  {year} {2014})}\BibitemShut {NoStop}%
\bibitem [{\citenamefont {Yin}\ \emph {et~al.}(2013{\natexlab{a}})\citenamefont
  {Yin}, \citenamefont {Liu}, \citenamefont {Tsvelik}, \citenamefont {Dean},
  \citenamefont {Upton}, \citenamefont {Kim}, \citenamefont {Casa},
  \citenamefont {Said}, \citenamefont {Gog}, \citenamefont {Qi}, \citenamefont
  {Cao},\ and\ \citenamefont {Hill}}]{Yin2013}%
  \BibitemOpen
  \bibfield  {author} {\bibinfo {author} {\bibfnamefont {W.-G.}\ \bibnamefont
  {Yin}}, \bibinfo {author} {\bibfnamefont {X.}~\bibnamefont {Liu}}, \bibinfo
  {author} {\bibfnamefont {a.~M.}\ \bibnamefont {Tsvelik}}, \bibinfo {author}
  {\bibfnamefont {M.~P.~M.}\ \bibnamefont {Dean}}, \bibinfo {author}
  {\bibfnamefont {M.~H.}\ \bibnamefont {Upton}}, \bibinfo {author}
  {\bibfnamefont {J.}~\bibnamefont {Kim}}, \bibinfo {author} {\bibfnamefont
  {D.}~\bibnamefont {Casa}}, \bibinfo {author} {\bibfnamefont {A.}~\bibnamefont
  {Said}}, \bibinfo {author} {\bibfnamefont {T.}~\bibnamefont {Gog}}, \bibinfo
  {author} {\bibfnamefont {T.~F.}\ \bibnamefont {Qi}}, \bibinfo {author}
  {\bibfnamefont {G.}~\bibnamefont {Cao}}, \ and\ \bibinfo {author}
  {\bibfnamefont {J.~P.}\ \bibnamefont {Hill}},\ }\href {\doibase
  10.1103/PhysRevLett.111.057202} {\bibfield  {journal} {\bibinfo  {journal}
  {Phys. Rev. Lett.}\ }\textbf {\bibinfo {volume} {111}},\ \bibinfo {pages}
  {057202} (\bibinfo {year} {2013}{\natexlab{a}})}\BibitemShut {NoStop}%
\bibitem [{\citenamefont {Jain}\ \emph {et~al.}(2013)\citenamefont {Jain},
  \citenamefont {Portnichenko}, \citenamefont {Jang}, \citenamefont {Jackeli},
  \citenamefont {Friemel}, \citenamefont {Ivanov}, \citenamefont {Piovano},
  \citenamefont {Yusuf}, \citenamefont {Keimer},\ and\ \citenamefont
  {Inosov}}]{Jain2013}%
  \BibitemOpen
  \bibfield  {author} {\bibinfo {author} {\bibfnamefont {A.}~\bibnamefont
  {Jain}}, \bibinfo {author} {\bibfnamefont {P.~Y.}\ \bibnamefont
  {Portnichenko}}, \bibinfo {author} {\bibfnamefont {H.}~\bibnamefont {Jang}},
  \bibinfo {author} {\bibfnamefont {G.}~\bibnamefont {Jackeli}}, \bibinfo
  {author} {\bibfnamefont {G.}~\bibnamefont {Friemel}}, \bibinfo {author}
  {\bibfnamefont {A.}~\bibnamefont {Ivanov}}, \bibinfo {author} {\bibfnamefont
  {A.}~\bibnamefont {Piovano}}, \bibinfo {author} {\bibfnamefont {S.~M.}\
  \bibnamefont {Yusuf}}, \bibinfo {author} {\bibfnamefont {B.}~\bibnamefont
  {Keimer}}, \ and\ \bibinfo {author} {\bibfnamefont {D.~S.}\ \bibnamefont
  {Inosov}},\ }\href {\doibase 10.1103/PhysRevB.88.224403} {\bibfield
  {journal} {\bibinfo  {journal} {Phys. Rev. B}\
  }\textbf {\bibinfo {volume} {88}},\ \bibinfo {pages} {224403} (\bibinfo
  {year} {2013})}\BibitemShut {NoStop}%
\bibitem [{\citenamefont {Agrestini}\ \emph {et~al.}(2014)\citenamefont
  {Agrestini}, \citenamefont {Adroja}, \citenamefont {Rotter}, \citenamefont
  {Majumdar}, \citenamefont {Lees}, \citenamefont {{Balakrishnan G.}},
  \citenamefont {Paul},\ and\ \citenamefont {Yeung}}]{Agrestini2014}%
  \BibitemOpen
  \bibfield  {author} {\bibinfo {author} {\bibfnamefont {S.}~\bibnamefont
  {Agrestini}}, \bibinfo {author} {\bibfnamefont {D.~T.}\ \bibnamefont
  {Adroja}}, \bibinfo {author} {\bibfnamefont {M.}~\bibnamefont {Rotter}},
  \bibinfo {author} {\bibfnamefont {S.}~\bibnamefont {Majumdar}}, \bibinfo
  {author} {\bibfnamefont {M.~R.}\ \bibnamefont {Lees}}, \bibinfo {author}
  {\bibnamefont {{Balakrishnan G.}}}, \bibinfo {author} {\bibfnamefont
  {D.}~\bibnamefont {Paul}}, \ and\ \bibinfo {author} {\bibfnamefont {Y.~Y.}\
  \bibnamefont {Yeung}},\ }\href@noop {} {\bibfield  {journal} {\bibinfo
  {journal} {unpublished}\ } (\bibinfo {year} {2014})}\BibitemShut {NoStop}%
\bibitem [{\citenamefont {Agrestini}\ \emph {et~al.}(2008)\citenamefont
  {Agrestini}, \citenamefont {Chapon}, \citenamefont {Daoud-Aladine},
  \citenamefont {Schefer}, \citenamefont {Gukasov}, \citenamefont {Mazzoli},
  \citenamefont {Lees},\ and\ \citenamefont {Petrenko}}]{Agrestini2008}%
  \BibitemOpen
  \bibfield  {author} {\bibinfo {author} {\bibfnamefont {S.}~\bibnamefont
  {Agrestini}}, \bibinfo {author} {\bibfnamefont {L.~C.}\ \bibnamefont
  {Chapon}}, \bibinfo {author} {\bibfnamefont {A.}~\bibnamefont
  {Daoud-Aladine}}, \bibinfo {author} {\bibfnamefont {J.}~\bibnamefont
  {Schefer}}, \bibinfo {author} {\bibfnamefont {A.}~\bibnamefont {Gukasov}},
  \bibinfo {author} {\bibfnamefont {C.}~\bibnamefont {Mazzoli}}, \bibinfo
  {author} {\bibfnamefont {M.~R.}\ \bibnamefont {Lees}}, \ and\ \bibinfo
  {author} {\bibfnamefont {O.~A.}\ \bibnamefont {Petrenko}},\ }\href {\doibase
  10.1103/PhysRevLett.101.097207} {\bibfield  {journal} {\bibinfo  {journal}
  {Phys. Rev. Lett.}\ }\textbf {\bibinfo {volume} {101}},\ \bibinfo {pages}
  {097207} (\bibinfo {year} {2008})}\BibitemShut {NoStop}%
\bibitem [{\citenamefont {Wu}\ \emph {et~al.}(2005)\citenamefont {Wu},
  \citenamefont {Haverkort}, \citenamefont {Hu}, \citenamefont {Khomskii},\
  and\ \citenamefont {Tjeng}}]{Wu2005}%
  \BibitemOpen
  \bibfield  {author} {\bibinfo {author} {\bibfnamefont {H.}~\bibnamefont
  {Wu}}, \bibinfo {author} {\bibfnamefont {M.~W.}\ \bibnamefont {Haverkort}},
  \bibinfo {author} {\bibfnamefont {Z.}~\bibnamefont {Hu}}, \bibinfo {author}
  {\bibfnamefont {D.~I.}\ \bibnamefont {Khomskii}}, \ and\ \bibinfo {author}
  {\bibfnamefont {L.~H.}\ \bibnamefont {Tjeng}},\ }\href {\doibase
  10.1103/PhysRevLett.95.186401} {\bibfield  {journal} {\bibinfo  {journal}
  {Phys. Rev. Lett.}\ }\textbf {\bibinfo {volume} {95}},\ \bibinfo {pages}
  {186401} (\bibinfo {year} {2005})}\BibitemShut {NoStop}%
\bibitem [{\citenamefont {Chapon}(2009)}]{Chapon2009}%
  \BibitemOpen
  \bibfield  {author} {\bibinfo {author} {\bibfnamefont {L.~C.}\ \bibnamefont
  {Chapon}},\ }\href {\doibase 10.1103/PhysRevB.80.172405} {\bibfield
  {journal} {\bibinfo  {journal} {Phys. Rev. B}\
  }\textbf {\bibinfo {volume} {80}},\ \bibinfo {pages} {172405} (\bibinfo
  {year} {2009})}\BibitemShut {NoStop}%
\bibitem [{\citenamefont {Sampathkumaran}\ and\ \citenamefont
  {Niazi}(2002)}]{Sampathkumaran2002}%
  \BibitemOpen
  \bibfield  {author} {\bibinfo {author} {\bibfnamefont {E.~V.}\ \bibnamefont
  {Sampathkumaran}}\ and\ \bibinfo {author} {\bibfnamefont {A.}~\bibnamefont
  {Niazi}},\ }\href {\doibase 10.1103/PhysRevB.65.180401} {\bibfield  {journal}
  {\bibinfo  {journal} {Phys. Rev. B}\ }\textbf {\bibinfo {volume} {65}},\
  \bibinfo {pages} {180401(R)} (\bibinfo {year} {2002})}\BibitemShut {NoStop}%
\bibitem [{\citenamefont {Sampathkumaran}\ \emph
  {et~al.}(2004{\natexlab{b}})\citenamefont {Sampathkumaran}, \citenamefont
  {Hiroi}, \citenamefont {Rayaprol},\ and\ \citenamefont
  {Uwatoko}}]{Sampathkumaran2004a}%
  \BibitemOpen
  \bibfield  {author} {\bibinfo {author} {\bibfnamefont {E.~V.}\ \bibnamefont
  {Sampathkumaran}}, \bibinfo {author} {\bibfnamefont {Z.}~\bibnamefont
  {Hiroi}}, \bibinfo {author} {\bibfnamefont {S.}~\bibnamefont {Rayaprol}}, \
  and\ \bibinfo {author} {\bibfnamefont {Y.}~\bibnamefont {Uwatoko}},\ }\href
  {\doibase 10.1016/j.jmmm.2004.07.028} {\bibfield  {journal} {\bibinfo
  {journal} {J. Magn. Magn. Mater.}\ }\textbf {\bibinfo {volume} {284}},\
  \bibinfo {pages} {L7} (\bibinfo {year} {2004}{\natexlab{b}})}\BibitemShut
  {NoStop}%
\bibitem [{\citenamefont {Sampathkumaran}\ \emph {et~al.}(2007)\citenamefont
  {Sampathkumaran}, \citenamefont {Mohapatra}, \citenamefont {Rayaprol},\ and\
  \citenamefont {Iyer}}]{Sampathkumaran2007}%
  \BibitemOpen
  \bibfield  {author} {\bibinfo {author} {\bibfnamefont {E.~V.}\ \bibnamefont
  {Sampathkumaran}}, \bibinfo {author} {\bibfnamefont {N.}~\bibnamefont
  {Mohapatra}}, \bibinfo {author} {\bibfnamefont {S.}~\bibnamefont {Rayaprol}},
  \ and\ \bibinfo {author} {\bibfnamefont {K.~K.}\ \bibnamefont {Iyer}},\
  }\href {\doibase 10.1103/PhysRevB.75.052412} {\bibfield  {journal} {\bibinfo
  {journal} {Phys. Rev. B}\ }\textbf {\bibinfo {volume} {75}},\ \bibinfo
  {pages} {052412} (\bibinfo {year} {2007})}\BibitemShut {NoStop}%
\bibitem [{\citenamefont {Paulose}\ \emph {et~al.}(2008)\citenamefont
  {Paulose}, \citenamefont {Mohapatra},\ and\ \citenamefont
  {Sampathkumaran}}]{Paulose2008}%
  \BibitemOpen
  \bibfield  {author} {\bibinfo {author} {\bibfnamefont {P.~L.}\ \bibnamefont
  {Paulose}}, \bibinfo {author} {\bibfnamefont {N.}~\bibnamefont {Mohapatra}},
  \ and\ \bibinfo {author} {\bibfnamefont {E.~V.}\ \bibnamefont
  {Sampathkumaran}},\ }\href {\doibase 10.1103/PhysRevB.77.172403} {\bibfield
  {journal} {\bibinfo  {journal} {Phys. Rev. B}\ }\textbf {\bibinfo {volume}
  {77}},\ \bibinfo {pages} {172403} (\bibinfo {year} {2008})}\BibitemShut
  {NoStop}%
\bibitem [{\citenamefont {Bindu}\ \emph {et~al.}(2009)\citenamefont {Bindu},
  \citenamefont {Maiti}, \citenamefont {Khalid},\ and\ \citenamefont
  {Sampathkumaran}}]{Bindu2009}%
  \BibitemOpen
  \bibfield  {author} {\bibinfo {author} {\bibfnamefont {R.}~\bibnamefont
  {Bindu}}, \bibinfo {author} {\bibfnamefont {K.}~\bibnamefont {Maiti}},
  \bibinfo {author} {\bibfnamefont {S.}~\bibnamefont {Khalid}}, \ and\ \bibinfo
  {author} {\bibfnamefont {E.~V.}\ \bibnamefont {Sampathkumaran}},\ }\href
  {\doibase 10.1103/PhysRevB.79.094103} {\bibfield  {journal} {\bibinfo
  {journal} {Phys. Rev. B}\ }\textbf {\bibinfo
  {volume} {79}},\ \bibinfo {pages} {094103} (\bibinfo {year}
  {2009})}\BibitemShut {NoStop}%
\bibitem [{\citenamefont {Gohil}\ \emph {et~al.}(2010)\citenamefont {Gohil},
  \citenamefont {Iyer}, \citenamefont {Aswathi}, \citenamefont {Ghosh},\ and\
  \citenamefont {Sampathkumaran}}]{Gohil2010}%
  \BibitemOpen
  \bibfield  {author} {\bibinfo {author} {\bibfnamefont {S.}~\bibnamefont
  {Gohil}}, \bibinfo {author} {\bibfnamefont {K.~K.}\ \bibnamefont {Iyer}},
  \bibinfo {author} {\bibfnamefont {P.}~\bibnamefont {Aswathi}}, \bibinfo
  {author} {\bibfnamefont {S.}~\bibnamefont {Ghosh}}, \ and\ \bibinfo {author}
  {\bibfnamefont {E.~V.}\ \bibnamefont {Sampathkumaran}},\ }\href {\doibase
  10.1063/1.3512899} {\bibfield  {journal} {\bibinfo  {journal} {J. Appl.
  Phys.}\ }\textbf {\bibinfo {volume} {108}},\ \bibinfo {pages} {103517}
  (\bibinfo {year} {2010})}\BibitemShut {NoStop}%
\bibitem [{\citenamefont {Niazi}\ \emph
  {et~al.}(2002{\natexlab{a}})\citenamefont {Niazi}, \citenamefont
  {Sampathkumaran}, \citenamefont {Paulose}, \citenamefont {Eckert},
  \citenamefont {Handstein},\ and\ \citenamefont {M\"{u}ller}}]{Niazi2002}%
  \BibitemOpen
  \bibfield  {author} {\bibinfo {author} {\bibfnamefont {A.}~\bibnamefont
  {Niazi}}, \bibinfo {author} {\bibfnamefont {E.~V.}\ \bibnamefont
  {Sampathkumaran}}, \bibinfo {author} {\bibfnamefont {P.~L.}\ \bibnamefont
  {Paulose}}, \bibinfo {author} {\bibfnamefont {D.}~\bibnamefont {Eckert}},
  \bibinfo {author} {\bibfnamefont {A.}~\bibnamefont {Handstein}}, \ and\
  \bibinfo {author} {\bibfnamefont {K.~H.}\ \bibnamefont {M\"{u}ller}},\ }\href
  {\doibase 10.1103/PhysRevB.65.064418} {\bibfield  {journal} {\bibinfo
  {journal} {Phys. Rev. B}\ }\textbf {\bibinfo {volume} {65}},\ \bibinfo
  {pages} {064418} (\bibinfo {year} {2002}{\natexlab{a}})}\BibitemShut
  {NoStop}%
\bibitem [{\citenamefont {Rayaprol}\ \emph {et~al.}(2004)\citenamefont
  {Rayaprol}, \citenamefont {Sengupta}, \citenamefont {Sampathkumaran},\ and\
  \citenamefont {Matsushita}}]{Rayaprol2004}%
  \BibitemOpen
  \bibfield  {author} {\bibinfo {author} {\bibfnamefont {S.}~\bibnamefont
  {Rayaprol}}, \bibinfo {author} {\bibfnamefont {K.}~\bibnamefont {Sengupta}},
  \bibinfo {author} {\bibfnamefont {E.~V.}\ \bibnamefont {Sampathkumaran}}, \
  and\ \bibinfo {author} {\bibfnamefont {Y.}~\bibnamefont {Matsushita}},\
  }\href {\doibase 10.1016/j.jssc.2004.05.050} {\bibfield  {journal} {\bibinfo
  {journal} {J. Solid State Chem.}\ }\textbf {\bibinfo {volume} {177}},\
  \bibinfo {pages} {3270} (\bibinfo {year} {2004})}\BibitemShut {NoStop}%
\bibitem [{\citenamefont {Mohapatra}\ \emph {et~al.}(2007)\citenamefont
  {Mohapatra}, \citenamefont {Iyer}, \citenamefont {Rayaprol},\ and\
  \citenamefont {Sampathkumaran}}]{Mohapatra2007}%
  \BibitemOpen
  \bibfield  {author} {\bibinfo {author} {\bibfnamefont {N.}~\bibnamefont
  {Mohapatra}}, \bibinfo {author} {\bibfnamefont {K.~K.}\ \bibnamefont {Iyer}},
  \bibinfo {author} {\bibfnamefont {S.}~\bibnamefont {Rayaprol}}, \ and\
  \bibinfo {author} {\bibfnamefont {E.~V.}\ \bibnamefont {Sampathkumaran}},\
  }\href {\doibase 10.1103/PhysRevB.75.214422} {\bibfield  {journal} {\bibinfo
  {journal} {Phys. Rev. B}\ }\textbf {\bibinfo {volume} {75}},\ \bibinfo
  {pages} {214422} (\bibinfo {year} {2007})}\BibitemShut {NoStop}%
\bibitem [{\citenamefont {Nguyen}(1994)}]{Nguyen1994}%
  \BibitemOpen
  \bibfield  {author} {\bibinfo {author} {\bibfnamefont {T.~N.}\ \bibnamefont
  {Nguyen}},\ }\emph {\bibinfo {title} {{Electrosynthesis and Characterisation
  of Main group and Transition Metal Oxides}}},\ \href
  {https://dspace.mit.edu/bitstream/handle/1721.1/11949/31053584.pdf?sequence=1}
  {\bibinfo {type} {PhD thesis}},\ \bibinfo  {school} {MIT} (\bibinfo {year}
  {1994})\BibitemShut {NoStop}%
\bibitem [{\citenamefont {Flahaut}\ \emph
  {et~al.}(2003{\natexlab{b}})\citenamefont {Flahaut}, \citenamefont {Hebert},
  \citenamefont {Maignan}, \citenamefont {Hardy}, \citenamefont {Martin},
  \citenamefont {Hervieu}, \citenamefont {Costes}, \citenamefont {Raquet},\
  and\ \citenamefont {Broto}}]{Flahaut2003a}%
  \BibitemOpen
  \bibfield  {author} {\bibinfo {author} {\bibfnamefont {D.}~\bibnamefont
  {Flahaut}}, \bibinfo {author} {\bibfnamefont {S.}~\bibnamefont {Hebert}},
  \bibinfo {author} {\bibfnamefont {A.}~\bibnamefont {Maignan}}, \bibinfo
  {author} {\bibfnamefont {V.}~\bibnamefont {Hardy}}, \bibinfo {author}
  {\bibfnamefont {C.}~\bibnamefont {Martin}}, \bibinfo {author} {\bibfnamefont
  {M.}~\bibnamefont {Hervieu}}, \bibinfo {author} {\bibfnamefont
  {M.}~\bibnamefont {Costes}}, \bibinfo {author} {\bibfnamefont
  {B.}~\bibnamefont {Raquet}}, \ and\ \bibinfo {author} {\bibfnamefont {J.~M.}\
  \bibnamefont {Broto}},\ }\href {\doibase 10.1140/epjb/e2003-00283-3}
  {\bibfield  {journal} {\bibinfo  {journal} {Eur. Phys. J. B
  }\ }\textbf {\bibinfo {volume} {35}},\ \bibinfo {pages} {317} (\bibinfo
  {year} {2003}{\natexlab{b}})}\BibitemShut {NoStop}%
\bibitem [{\citenamefont {Mikhailova}\ \emph
  {et~al.}(2012{\natexlab{b}})\citenamefont {Mikhailova}, \citenamefont
  {Schwarz}, \citenamefont {Senyshyn}, \citenamefont {Bell}, \citenamefont
  {Skourski}, \citenamefont {Ehrenberg}, \citenamefont {Tsirlin}, \citenamefont
  {Agrestini}, \citenamefont {Rotter}, \citenamefont {Reichel}, \citenamefont
  {Chen}, \citenamefont {Hu}, \citenamefont {Li}, \citenamefont {Li},\ and\
  \citenamefont {Tjeng}}]{Mikhailova2012a}%
  \BibitemOpen
  \bibfield  {author} {\bibinfo {author} {\bibfnamefont {D.}~\bibnamefont
  {Mikhailova}}, \bibinfo {author} {\bibfnamefont {B.}~\bibnamefont {Schwarz}},
  \bibinfo {author} {\bibfnamefont {A.}~\bibnamefont {Senyshyn}}, \bibinfo
  {author} {\bibfnamefont {A.~M.~T.}\ \bibnamefont {Bell}}, \bibinfo {author}
  {\bibfnamefont {Y.}~\bibnamefont {Skourski}}, \bibinfo {author}
  {\bibfnamefont {H.}~\bibnamefont {Ehrenberg}}, \bibinfo {author}
  {\bibfnamefont {A.~A.}\ \bibnamefont {Tsirlin}}, \bibinfo {author}
  {\bibfnamefont {S.}~\bibnamefont {Agrestini}}, \bibinfo {author}
  {\bibfnamefont {M.}~\bibnamefont {Rotter}}, \bibinfo {author} {\bibfnamefont
  {P.}~\bibnamefont {Reichel}}, \bibinfo {author} {\bibfnamefont {J.~M.}\
  \bibnamefont {Chen}}, \bibinfo {author} {\bibfnamefont {Z.}~\bibnamefont
  {Hu}}, \bibinfo {author} {\bibfnamefont {Z.~M.}\ \bibnamefont {Li}}, \bibinfo
  {author} {\bibfnamefont {Z.~F.}\ \bibnamefont {Li}}, \ and\ \bibinfo {author}
  {\bibfnamefont {L.~H.}\ \bibnamefont {Tjeng}},\ }\href {\doibase
  10.1103/PhysRevB.86.134409} {\bibfield  {journal} {\bibinfo  {journal} {Phys.
  Rev. B}\ }\textbf {\bibinfo {volume} {86}},\
  \bibinfo {pages} {134409} (\bibinfo {year} {2012}{\natexlab{b}})}\BibitemShut
  {NoStop}%
\bibitem [{\citenamefont {Niazi}\ \emph
  {et~al.}(2002{\natexlab{b}})\citenamefont {Niazi}, \citenamefont {Paulose},\
  and\ \citenamefont {Sampathkumaran}}]{Niazi2002a}%
  \BibitemOpen
  \bibfield  {author} {\bibinfo {author} {\bibfnamefont {A.}~\bibnamefont
  {Niazi}}, \bibinfo {author} {\bibfnamefont {P.~L.}\ \bibnamefont {Paulose}},
  \ and\ \bibinfo {author} {\bibfnamefont {E.~V.}\ \bibnamefont
  {Sampathkumaran}},\ }\href {\doibase 10.1103/PhysRevLett.88.107202}
  {\bibfield  {journal} {\bibinfo  {journal} {Phys. Rev. Lett.}\ }\textbf
  {\bibinfo {volume} {88}},\ \bibinfo {pages} {107202} (\bibinfo {year}
  {2002}{\natexlab{b}})}\BibitemShut {NoStop}%
\bibitem [{\citenamefont {Niazi}\ \emph {et~al.}(2001)\citenamefont {Niazi},
  \citenamefont {Sampathkumaran}, \citenamefont {Paulose}, \citenamefont
  {Eckert}, \citenamefont {Handstein},\ and\ \citenamefont
  {M\"{u}ller}}]{Niazi2001}%
  \BibitemOpen
  \bibfield  {author} {\bibinfo {author} {\bibfnamefont {A.}~\bibnamefont
  {Niazi}}, \bibinfo {author} {\bibfnamefont {E.~V.}\ \bibnamefont
  {Sampathkumaran}}, \bibinfo {author} {\bibfnamefont {P.~L.}\ \bibnamefont
  {Paulose}}, \bibinfo {author} {\bibfnamefont {D.}~\bibnamefont {Eckert}},
  \bibinfo {author} {\bibfnamefont {A.}~\bibnamefont {Handstein}}, \ and\
  \bibinfo {author} {\bibfnamefont {K.~H.}\ \bibnamefont {M\"{u}ller}},\ }\href
  {\doibase 10.1016/S0038-1098(01)00313-1} {\bibfield  {journal} {\bibinfo
  {journal} {Solid State Commun.}\ }\textbf {\bibinfo {volume} {120}},\
  \bibinfo {pages} {11} (\bibinfo {year} {2001})}\BibitemShut {NoStop}%
\bibitem [{\citenamefont {Yin}\ \emph {et~al.}(2013{\natexlab{b}})\citenamefont
  {Yin}, \citenamefont {Liu}, \citenamefont {Tsvelik}, \citenamefont {Dean},
  \citenamefont {Upton}, \citenamefont {Kim}, \citenamefont {Casa},
  \citenamefont {Said}, \citenamefont {Gog}, \citenamefont {Qi}, \citenamefont
  {Cao},\ and\ \citenamefont {Hill}}]{Yin2013a}%
  \BibitemOpen
  \bibfield  {author} {\bibinfo {author} {\bibfnamefont {W.~G.}\ \bibnamefont
  {Yin}}, \bibinfo {author} {\bibfnamefont {X.}~\bibnamefont {Liu}}, \bibinfo
  {author} {\bibfnamefont {A.~M.}\ \bibnamefont {Tsvelik}}, \bibinfo {author}
  {\bibfnamefont {M.~P.~M.}\ \bibnamefont {Dean}}, \bibinfo {author}
  {\bibfnamefont {M.~H.}\ \bibnamefont {Upton}}, \bibinfo {author}
  {\bibfnamefont {J.}~\bibnamefont {Kim}}, \bibinfo {author} {\bibfnamefont
  {D.}~\bibnamefont {Casa}}, \bibinfo {author} {\bibfnamefont {A.}~\bibnamefont
  {Said}}, \bibinfo {author} {\bibfnamefont {T.}~\bibnamefont {Gog}}, \bibinfo
  {author} {\bibfnamefont {T.~F.}\ \bibnamefont {Qi}}, \bibinfo {author}
  {\bibfnamefont {G.}~\bibnamefont {Cao}}, \ and\ \bibinfo {author}
  {\bibfnamefont {J.~P.}\ \bibnamefont {Hill}},\ }\href {\doibase
  10.1103/PhysRevLett.111.057202} {\bibfield  {journal} {\bibinfo  {journal}
  {Phys. Rev. Lett.}\ }\textbf {\bibinfo {volume} {111}},\ \bibinfo {pages}
  {057202} (\bibinfo {year} {2013}{\natexlab{b}})}\BibitemShut {NoStop}%
\bibitem [{\citenamefont {Nguyen}\ and\ \citenamefont {zur
  Loye}(1995{\natexlab{b}})}]{Nguyen1995a}%
  \BibitemOpen
  \bibfield  {author} {\bibinfo {author} {\bibfnamefont {T.}~\bibnamefont
  {Nguyen}}\ and\ \bibinfo {author} {\bibfnamefont {H.-C.}\ \bibnamefont {zur
  Loye}},\ }\href {\doibase 10.1006/jssc.1995.1277} {\bibfield  {journal}
  {\bibinfo  {journal} {J. Solid State Chem.}\ }\textbf {\bibinfo {volume}
  {117}},\ \bibinfo {pages} {300} (\bibinfo {year}
  {1995}{\natexlab{b}})}\BibitemShut {NoStop}%
\bibitem [{\citenamefont {Lefran\c{c}ois}\ \emph {et~al.}(2014)\citenamefont
  {Lefran\c{c}ois}, \citenamefont {Chapon}, \citenamefont {Simonet},
  \citenamefont {Lejay}, \citenamefont {Khalyavin}, \citenamefont {Rayaprol},
  \citenamefont {Sampathkumaran}, \citenamefont {Ballou},\ and\ \citenamefont
  {Adroja}}]{Lefrancois2014a}%
  \BibitemOpen
  \bibfield  {author} {\bibinfo {author} {\bibfnamefont {E.}~\bibnamefont
  {Lefran\c{c}ois}}, \bibinfo {author} {\bibfnamefont {L.~C.}\ \bibnamefont
  {Chapon}}, \bibinfo {author} {\bibfnamefont {V.}~\bibnamefont {Simonet}},
  \bibinfo {author} {\bibfnamefont {P.}~\bibnamefont {Lejay}}, \bibinfo
  {author} {\bibfnamefont {D.}~\bibnamefont {Khalyavin}}, \bibinfo {author}
  {\bibfnamefont {S.}~\bibnamefont {Rayaprol}}, \bibinfo {author}
  {\bibfnamefont {E.~V.}\ \bibnamefont {Sampathkumaran}}, \bibinfo {author}
  {\bibfnamefont {R.}~\bibnamefont {Ballou}}, \ and\ \bibinfo {author}
  {\bibfnamefont {D.~T.}\ \bibnamefont {Adroja}},\ }\href {\doibase
  10.1103/PhysRevB.90.014408} {\bibfield  {journal} {\bibinfo  {journal} {Phys.
  Rev. B}\ }\textbf {\bibinfo {volume} {90}},\
  \bibinfo {pages} {014408} (\bibinfo {year} {2014})}\BibitemShut {NoStop}%
\bibitem [{Sup()}]{Supp}%
  \BibitemOpen
  \href@noop {} {\ }\BibitemShut {NoStop}%
\bibitem [{\citenamefont {Bloch}(1930)}]{Bloch1930a}%
  \BibitemOpen
  \bibfield  {author} {\bibinfo {author} {\bibfnamefont {F.}~\bibnamefont
  {Bloch}},\ }\href {\doibase 10.1007/BF01339661} {\bibfield  {journal}
  {\bibinfo  {journal} {Z. Phys.}\ }\textbf {\bibinfo {volume}
  {61}},\ \bibinfo {pages} {206} (\bibinfo {year} {1930})}\BibitemShut
  {NoStop}%
\bibitem [{\citenamefont {Slater}(1930)}]{Slater1930}%
  \BibitemOpen
  \bibfield  {author} {\bibinfo {author} {\bibfnamefont {J.~C.}\ \bibnamefont
  {Slater}},\ }\href {\doibase 10.1103/PhysRev.35.509} {\bibfield  {journal}
  {\bibinfo  {journal} {Phys. Rev.}\ }\textbf {\bibinfo {volume} {35}},\
  \bibinfo {pages} {509} (\bibinfo {year} {1930})}\BibitemShut {NoStop}%
\bibitem [{\citenamefont {Toth}\ and\ \citenamefont {Lake}(2014)}]{Toth2014}%
  \BibitemOpen
  \bibfield  {author} {\bibinfo {author} {\bibfnamefont {S.}~\bibnamefont
  {Toth}}\ and\ \bibinfo {author} {\bibfnamefont {B.}~\bibnamefont {Lake}},\
  }\href {http://arxiv.org/abs/1402.6069} {\ \bibinfo {pages} {SpinW library, www.psi.ch/spinw} (\bibinfo
  {year} {2014})},\ \Eprint {http://arxiv.org/abs/1402.6069} {arXiv:1402.6069}
  \BibitemShut {NoStop}%
\bibitem [{\citenamefont {Lovesey}(1986)}]{Lovesey1986}%
  \BibitemOpen
  \bibfield  {author} {\bibinfo {author} {\bibfnamefont {S.~W.}\ \bibnamefont
  {Lovesey}},\ }\href@noop {} {\emph {\bibinfo {title} {{The Theory of Neutron
  Scattering from Condensed Matter}}}}\ (\bibinfo {year} {1986})\BibitemShut
  {NoStop}%

\end{thebibliography}

\begin{thebibliography}{19}%
\makeatletter
\providecommand \@ifxundefined [1]{%
 \@ifx{#1\undefined}
}%
\providecommand \@ifnum [1]{%
 \ifnum #1\expandafter \@firstoftwo
 \else \expandafter \@secondoftwo
 \fi
}%
\providecommand \@ifx [1]{%
 \ifx #1\expandafter \@firstoftwo
 \else \expandafter \@secondoftwo
 \fi
}%
\providecommand \natexlab [1]{#1}%
\providecommand \enquote  [1]{``#1''}%
\providecommand \bibnamefont  [1]{#1}%
\providecommand \bibfnamefont [1]{#1}%
\providecommand \citenamefont [1]{#1}%
\providecommand \href@noop [0]{\@secondoftwo}%
\providecommand \href [0]{\begingroup \@sanitize@url \@href}%
\providecommand \@href[1]{\@@startlink{#1}\@@href}%
\providecommand \@@href[1]{\endgroup#1\@@endlink}%
\providecommand \@sanitize@url [0]{\catcode `\\12\catcode `\$12\catcode
  `\&12\catcode `\#12\catcode `\^12\catcode `\_12\catcode `\%12\relax}%
\providecommand \@@startlink[1]{}%
\providecommand \@@endlink[0]{}%
\providecommand \url  [0]{\begingroup\@sanitize@url \@url }%
\providecommand \@url [1]{\endgroup\@href {#1}{\urlprefix }}%
\providecommand \urlprefix  [0]{URL }%
\providecommand \Eprint [0]{\href }%
\providecommand \doibase [0]{http://dx.doi.org/}%
\providecommand \selectlanguage [0]{\@gobble}%
\providecommand \bibinfo  [0]{\@secondoftwo}%
\providecommand \bibfield  [0]{\@secondoftwo}%
\providecommand \translation [1]{[#1]}%
\providecommand \BibitemOpen [0]{}%
\providecommand \bibitemStop [0]{}%
\providecommand \bibitemNoStop [0]{.\EOS\space}%
\providecommand \EOS [0]{\spacefactor3000\relax}%
\providecommand \BibitemShut  [1]{\csname bibitem#1\endcsname}%
\let\auto@bib@innerbib\@empty
\bibitem [{\citenamefont {Lefran\c{c}ois}\ \emph
  {et~al.}(2014{\natexlab{a}})\citenamefont {Lefran\c{c}ois}, \citenamefont
  {Chapon}, \citenamefont {Simonet}, \citenamefont {Lejay}, \citenamefont
  {Khalyavin}, \citenamefont {Rayaprol}, \citenamefont {Sampathkumaran},
  \citenamefont {Ballou},\ and\ \citenamefont {Adroja}}]{Lefrancois2014asm}%
  \BibitemOpen
  \bibfield  {author} {\bibinfo {author} {\bibfnamefont {E.}~\bibnamefont
  {Lefran\c{c}ois}}, \bibinfo {author} {\bibfnamefont {L.~C.}\ \bibnamefont
  {Chapon}}, \bibinfo {author} {\bibfnamefont {V.}~\bibnamefont {Simonet}},
  \bibinfo {author} {\bibfnamefont {P.}~\bibnamefont {Lejay}}, \bibinfo
  {author} {\bibfnamefont {D.}~\bibnamefont {Khalyavin}}, \bibinfo {author}
  {\bibfnamefont {S.}~\bibnamefont {Rayaprol}}, \bibinfo {author}
  {\bibfnamefont {E.~V.}\ \bibnamefont {Sampathkumaran}}, \bibinfo {author}
  {\bibfnamefont {R.}~\bibnamefont {Ballou}}, \ and\ \bibinfo {author}
  {\bibfnamefont {D.~T.}\ \bibnamefont {Adroja}},\ }\href {\doibase
  10.1103/PhysRevB.90.014408} {\bibfield  {journal} {\bibinfo  {journal} {Phys.
  Rev. B}\ }\textbf {\bibinfo {volume} {90}},\
  \bibinfo {pages} {014408} (\bibinfo {year} {2014}{\natexlab{a}})}\BibitemShut
  {NoStop}%
\bibitem [{\citenamefont {Lefran\c{c}ois}\ \emph
  {et~al.}(2014{\natexlab{b}})\citenamefont {Lefran\c{c}ois}, \citenamefont
  {Chapon}, \citenamefont {Simonet}, \citenamefont {Lejay}, \citenamefont
  {Khalyavin}, \citenamefont {Rayaprol}, \citenamefont {Sampathkumaran},
  \citenamefont {Ballou},\ and\ \citenamefont {Adroja}}]{Lefrancois2014sm}%
  \BibitemOpen
  \bibfield  {author} {\bibinfo {author} {\bibfnamefont {E.}~\bibnamefont
  {Lefran\c{c}ois}}, \bibinfo {author} {\bibfnamefont {L.~C.}\ \bibnamefont
  {Chapon}}, \bibinfo {author} {\bibfnamefont {V.}~\bibnamefont {Simonet}},
  \bibinfo {author} {\bibfnamefont {P.}~\bibnamefont {Lejay}}, \bibinfo
  {author} {\bibfnamefont {D.}~\bibnamefont {Khalyavin}}, \bibinfo {author}
  {\bibfnamefont {S.}~\bibnamefont {Rayaprol}}, \bibinfo {author}
  {\bibfnamefont {E.~V.}\ \bibnamefont {Sampathkumaran}}, \bibinfo {author}
  {\bibfnamefont {R.}~\bibnamefont {Ballou}}, \ and\ \bibinfo {author}
  {\bibfnamefont {D.~T.}\ \bibnamefont {Adroja}},\ }\href {\doibase
  10.1103/PhysRevB.90.014408} {\bibfield  {journal} {\bibinfo  {journal} {Phys.
  Rev. B}\ }\textbf {\bibinfo {volume} {90}},\ \bibinfo {pages} {014408}
  (\bibinfo {year} {2014}{\natexlab{b}})}\BibitemShut {NoStop}%
\bibitem [{\citenamefont {Nguyen}\ and\ \citenamefont {zur
  Loye}(1995{\natexlab{a}})}]{Nguyen1995sm}%
  \BibitemOpen
  \bibfield  {author} {\bibinfo {author} {\bibfnamefont {T.}~\bibnamefont
  {Nguyen}}\ and\ \bibinfo {author} {\bibfnamefont {H.-C.}\ \bibnamefont {zur
  Loye}},\ }\href {\doibase 10.1006/jssc.1995.1277} {\bibfield  {journal}
  {\bibinfo  {journal} {J. Solid State Chem.}\ }\textbf {\bibinfo {volume}
  {117}},\ \bibinfo {pages} {300} (\bibinfo {year}
  {1995}{\natexlab{a}})}\BibitemShut {NoStop}%
\bibitem [{\citenamefont {Vajenine}\ \emph {et~al.}(1996)\citenamefont
  {Vajenine}, \citenamefont {Hoffmann},\ and\ \citenamefont {{Zur
  Loye}}}]{Vajenine1996sm}%
  \BibitemOpen
  \bibfield  {author} {\bibinfo {author} {\bibfnamefont {G.~V.}\ \bibnamefont
  {Vajenine}}, \bibinfo {author} {\bibfnamefont {R.}~\bibnamefont {Hoffmann}},
  \ and\ \bibinfo {author} {\bibfnamefont {H.~C.}\ \bibnamefont {{Zur Loye}}},\
  }\href {\doibase 10.1016/0301-0104(95)00335-5} {\bibfield  {journal}
  {\bibinfo  {journal} {Chem. Phys.}\ }\textbf {\bibinfo {volume} {204}},\
  \bibinfo {pages} {469} (\bibinfo {year} {1996})}\BibitemShut {NoStop}%
\bibitem [{\citenamefont {Flahaut}\ \emph
  {et~al.}(2003{\natexlab{a}})\citenamefont {Flahaut}, \citenamefont {Hebert},
  \citenamefont {Maignan}, \citenamefont {Hardy}, \citenamefont {Martin},
  \citenamefont {Hervieu}, \citenamefont {Costes}, \citenamefont {Raquet},\
  and\ \citenamefont {Broto}}]{Flahaut2003sm}%
  \BibitemOpen
  \bibfield  {author} {\bibinfo {author} {\bibfnamefont {D.}~\bibnamefont
  {Flahaut}}, \bibinfo {author} {\bibfnamefont {S.}~\bibnamefont {Hebert}},
  \bibinfo {author} {\bibfnamefont {A.}~\bibnamefont {Maignan}}, \bibinfo
  {author} {\bibfnamefont {V.}~\bibnamefont {Hardy}}, \bibinfo {author}
  {\bibfnamefont {C.}~\bibnamefont {Martin}}, \bibinfo {author} {\bibfnamefont
  {M.}~\bibnamefont {Hervieu}}, \bibinfo {author} {\bibfnamefont
  {M.}~\bibnamefont {Costes}}, \bibinfo {author} {\bibfnamefont
  {B.}~\bibnamefont {Raquet}}, \ and\ \bibinfo {author} {\bibfnamefont {J.~M.}\
  \bibnamefont {Broto}},\ }\href {\doibase 10.1140/epjb/e2003-00283-3}
  {\bibfield  {journal} {\bibinfo  {journal} {Eur. Phys. J. B
  }\ }\textbf {\bibinfo {volume} {35}},\ \bibinfo {pages} {317} (\bibinfo
  {year} {2003}{\natexlab{a}})}\BibitemShut {NoStop}%
\bibitem [{\citenamefont {Mikhailova}\ \emph
  {et~al.}(2012{\natexlab{a}})\citenamefont {Mikhailova}, \citenamefont
  {Schwarz}, \citenamefont {Senyshyn}, \citenamefont {Bell}, \citenamefont
  {Skourski}, \citenamefont {Ehrenberg}, \citenamefont {Tsirlin}, \citenamefont
  {Agrestini}, \citenamefont {Rotter}, \citenamefont {Reichel}, \citenamefont
  {Chen}, \citenamefont {Hu}, \citenamefont {Li}, \citenamefont {Li},\ and\
  \citenamefont {Tjeng}}]{Mikhailova2012sm}%
  \BibitemOpen
  \bibfield  {author} {\bibinfo {author} {\bibfnamefont {D.}~\bibnamefont
  {Mikhailova}}, \bibinfo {author} {\bibfnamefont {B.}~\bibnamefont {Schwarz}},
  \bibinfo {author} {\bibfnamefont {A.}~\bibnamefont {Senyshyn}}, \bibinfo
  {author} {\bibfnamefont {a.~M.~T.}\ \bibnamefont {Bell}}, \bibinfo {author}
  {\bibfnamefont {Y.}~\bibnamefont {Skourski}}, \bibinfo {author}
  {\bibfnamefont {H.}~\bibnamefont {Ehrenberg}}, \bibinfo {author}
  {\bibfnamefont {a.~a.}\ \bibnamefont {Tsirlin}}, \bibinfo {author}
  {\bibfnamefont {S.}~\bibnamefont {Agrestini}}, \bibinfo {author}
  {\bibfnamefont {M.}~\bibnamefont {Rotter}}, \bibinfo {author} {\bibfnamefont
  {P.}~\bibnamefont {Reichel}}, \bibinfo {author} {\bibfnamefont {J.~M.}\
  \bibnamefont {Chen}}, \bibinfo {author} {\bibfnamefont {Z.}~\bibnamefont
  {Hu}}, \bibinfo {author} {\bibfnamefont {Z.~M.}\ \bibnamefont {Li}}, \bibinfo
  {author} {\bibfnamefont {Z.~F.}\ \bibnamefont {Li}}, \ and\ \bibinfo {author}
  {\bibfnamefont {L.~H.}\ \bibnamefont {Tjeng}},\ }\href {\doibase
  10.1103/PhysRevB.86.134409} {\bibfield  {journal} {\bibinfo  {journal} {Phys.
  Rev. B}\ }\textbf {\bibinfo {volume} {86}},\ \bibinfo {pages} {134409}
  (\bibinfo {year} {2012}{\natexlab{a}})}\BibitemShut {NoStop}%
\bibitem [{\citenamefont {Flahaut}\ \emph
  {et~al.}(2003{\natexlab{b}})\citenamefont {Flahaut}, \citenamefont {Hebert},
  \citenamefont {Maignan}, \citenamefont {Hardy}, \citenamefont {Martin},
  \citenamefont {Hervieu}, \citenamefont {Costes}, \citenamefont {Raquet},\
  and\ \citenamefont {Broto}}]{Flahaut2003asm}%
  \BibitemOpen
  \bibfield  {author} {\bibinfo {author} {\bibfnamefont {D.}~\bibnamefont
  {Flahaut}}, \bibinfo {author} {\bibfnamefont {S.}~\bibnamefont {Hebert}},
  \bibinfo {author} {\bibfnamefont {A.}~\bibnamefont {Maignan}}, \bibinfo
  {author} {\bibfnamefont {V.}~\bibnamefont {Hardy}}, \bibinfo {author}
  {\bibfnamefont {C.}~\bibnamefont {Martin}}, \bibinfo {author} {\bibfnamefont
  {M.}~\bibnamefont {Hervieu}}, \bibinfo {author} {\bibfnamefont
  {M.}~\bibnamefont {Costes}}, \bibinfo {author} {\bibfnamefont
  {B.}~\bibnamefont {Raquet}}, \ and\ \bibinfo {author} {\bibfnamefont {J.~M.}\
  \bibnamefont {Broto}},\ }\href {\doibase 10.1140/epjb/e2003-00283-3}
  {\bibfield  {journal} {\bibinfo  {journal} {Eur. Phys. J. B
  }\ }\textbf {\bibinfo {volume} {35}},\ \bibinfo {pages} {317} (\bibinfo
  {year} {2003}{\natexlab{b}})}\BibitemShut {NoStop}%
\bibitem [{\citenamefont {Hillier}\ \emph {et~al.}(2011)\citenamefont
  {Hillier}, \citenamefont {Adroja}, \citenamefont {Kockelmann}, \citenamefont
  {Chapon}, \citenamefont {Rayaprol}, \citenamefont {Manuel}, \citenamefont
  {Michor},\ and\ \citenamefont {Sampathkumaran}}]{Hillier2011sm}%
  \BibitemOpen
  \bibfield  {author} {\bibinfo {author} {\bibfnamefont {A.~D.}\ \bibnamefont
  {Hillier}}, \bibinfo {author} {\bibfnamefont {D.~T.}\ \bibnamefont {Adroja}},
  \bibinfo {author} {\bibfnamefont {W.}~\bibnamefont {Kockelmann}}, \bibinfo
  {author} {\bibfnamefont {L.~C.}\ \bibnamefont {Chapon}}, \bibinfo {author}
  {\bibfnamefont {S.}~\bibnamefont {Rayaprol}}, \bibinfo {author}
  {\bibfnamefont {P.}~\bibnamefont {Manuel}}, \bibinfo {author} {\bibfnamefont
  {H.}~\bibnamefont {Michor}}, \ and\ \bibinfo {author} {\bibfnamefont {E.~V.}\
  \bibnamefont {Sampathkumaran}},\ }\href {\doibase 10.1103/PhysRevB.83.024414}
  {\bibfield  {journal} {\bibinfo  {journal} {Phys. Rev. B
  }\ }\textbf {\bibinfo {volume} {83}},\ \bibinfo {pages} {024414}
  (\bibinfo {year} {2011})}\BibitemShut {NoStop}%
\bibitem [{\citenamefont {Mikhailova}\ \emph
  {et~al.}(2012{\natexlab{b}})\citenamefont {Mikhailova}, \citenamefont
  {Schwarz}, \citenamefont {Senyshyn}, \citenamefont {Bell}, \citenamefont
  {Skourski}, \citenamefont {Ehrenberg}, \citenamefont {Tsirlin}, \citenamefont
  {Agrestini}, \citenamefont {Rotter}, \citenamefont {Reichel}, \citenamefont
  {Chen}, \citenamefont {Hu}, \citenamefont {Li}, \citenamefont {Li},\ and\
  \citenamefont {Tjeng}}]{Mikhailova2012asm}%
  \BibitemOpen
  \bibfield  {author} {\bibinfo {author} {\bibfnamefont {D.}~\bibnamefont
  {Mikhailova}}, \bibinfo {author} {\bibfnamefont {B.}~\bibnamefont {Schwarz}},
  \bibinfo {author} {\bibfnamefont {A.}~\bibnamefont {Senyshyn}}, \bibinfo
  {author} {\bibfnamefont {A.~M.~T.}\ \bibnamefont {Bell}}, \bibinfo {author}
  {\bibfnamefont {Y.}~\bibnamefont {Skourski}}, \bibinfo {author}
  {\bibfnamefont {H.}~\bibnamefont {Ehrenberg}}, \bibinfo {author}
  {\bibfnamefont {A.~A.}\ \bibnamefont {Tsirlin}}, \bibinfo {author}
  {\bibfnamefont {S.}~\bibnamefont {Agrestini}}, \bibinfo {author}
  {\bibfnamefont {M.}~\bibnamefont {Rotter}}, \bibinfo {author} {\bibfnamefont
  {P.}~\bibnamefont {Reichel}}, \bibinfo {author} {\bibfnamefont {J.~M.}\
  \bibnamefont {Chen}}, \bibinfo {author} {\bibfnamefont {Z.}~\bibnamefont
  {Hu}}, \bibinfo {author} {\bibfnamefont {Z.~M.}\ \bibnamefont {Li}}, \bibinfo
  {author} {\bibfnamefont {Z.~F.}\ \bibnamefont {Li}}, \ and\ \bibinfo {author}
  {\bibfnamefont {L.~H.}\ \bibnamefont {Tjeng}},\ }\href {\doibase
  10.1103/PhysRevB.86.134409} {\bibfield  {journal} {\bibinfo  {journal} {Phys.
  Rev. B}\ }\textbf {\bibinfo {volume} {86}},\
  \bibinfo {pages} {134409} (\bibinfo {year} {2012}{\natexlab{b}})}\BibitemShut
  {NoStop}%
\bibitem [{\citenamefont {Stitzer}\ \emph {et~al.}(2002)\citenamefont
  {Stitzer}, \citenamefont {Henley}, \citenamefont {Claridge}, \citenamefont
  {zur Loye},\ and\ \citenamefont {Layland}}]{Stitzer2002sm}%
  \BibitemOpen
  \bibfield  {author} {\bibinfo {author} {\bibfnamefont {K.}~\bibnamefont
  {Stitzer}}, \bibinfo {author} {\bibfnamefont {W.}~\bibnamefont {Henley}},
  \bibinfo {author} {\bibfnamefont {J.}~\bibnamefont {Claridge}}, \bibinfo
  {author} {\bibfnamefont {H.-C.}\ \bibnamefont {zur Loye}}, \ and\ \bibinfo
  {author} {\bibfnamefont {R.}~\bibnamefont {Layland}},\ }\href {\doibase
  10.1006/jssc.2001.9463} {\bibfield  {journal} {\bibinfo  {journal} {J. Solid
  State Chem.}\ }\textbf {\bibinfo {volume} {164}},\ \bibinfo {pages} {220}
  (\bibinfo {year} {2002})}\BibitemShut {NoStop}%
\bibitem [{\citenamefont {Rayaprol}\ \emph {et~al.}(2004)\citenamefont
  {Rayaprol}, \citenamefont {Sengupta}, \citenamefont {Sampathkumaran},\ and\
  \citenamefont {Matsushita}}]{Rayaprol2004sm}%
  \BibitemOpen
  \bibfield  {author} {\bibinfo {author} {\bibfnamefont {S.}~\bibnamefont
  {Rayaprol}}, \bibinfo {author} {\bibfnamefont {K.}~\bibnamefont {Sengupta}},
  \bibinfo {author} {\bibfnamefont {E.~V.}\ \bibnamefont {Sampathkumaran}}, \
  and\ \bibinfo {author} {\bibfnamefont {Y.}~\bibnamefont {Matsushita}},\
  }\href {\doibase 10.1016/j.jssc.2004.05.050} {\bibfield  {journal} {\bibinfo
  {journal} {J. Solid State Chem.}\ }\textbf {\bibinfo {volume} {177}},\
  \bibinfo {pages} {3270} (\bibinfo {year} {2004})}\BibitemShut {NoStop}%
\bibitem [{\citenamefont {Mohapatra}\ \emph {et~al.}(2007)\citenamefont
  {Mohapatra}, \citenamefont {Iyer}, \citenamefont {Rayaprol},\ and\
  \citenamefont {Sampathkumaran}}]{Mohapatra2007sm}%
  \BibitemOpen
  \bibfield  {author} {\bibinfo {author} {\bibfnamefont {N.}~\bibnamefont
  {Mohapatra}}, \bibinfo {author} {\bibfnamefont {K.~K.}\ \bibnamefont {Iyer}},
  \bibinfo {author} {\bibfnamefont {S.}~\bibnamefont {Rayaprol}}, \ and\
  \bibinfo {author} {\bibfnamefont {E.~V.}\ \bibnamefont {Sampathkumaran}},\
  }\href {\doibase 10.1103/PhysRevB.75.214422} {\bibfield  {journal} {\bibinfo
  {journal} {Phys. Rev. B}\ }\textbf {\bibinfo {volume} {75}},\ \bibinfo
  {pages} {214422} (\bibinfo {year} {2007})}\BibitemShut {NoStop}%
\bibitem [{\citenamefont {Nguyen}\ and\ \citenamefont {zur
  Loye}(1995{\natexlab{b}})}]{Nguyen1995asm}%
  \BibitemOpen
  \bibfield  {author} {\bibinfo {author} {\bibfnamefont {T.}~\bibnamefont
  {Nguyen}}\ and\ \bibinfo {author} {\bibfnamefont {H.-C.}\ \bibnamefont {zur
  Loye}},\ }\href {\doibase 10.1006/jssc.1995.1277} {\bibfield  {journal}
  {\bibinfo  {journal} {J. Solid State Chem.}\ }\textbf {\bibinfo {volume}
  {117}},\ \bibinfo {pages} {300} (\bibinfo {year}
  {1995}{\natexlab{b}})}\BibitemShut {NoStop}%
\bibitem [{\citenamefont {Niazi}\ \emph {et~al.}(2002)\citenamefont {Niazi},
  \citenamefont {Sampathkumaran}, \citenamefont {Paulose}, \citenamefont
  {Eckert}, \citenamefont {Handstein},\ and\ \citenamefont
  {M\"{u}ller}}]{Niazi2002sm}%
  \BibitemOpen
  \bibfield  {author} {\bibinfo {author} {\bibfnamefont {A.}~\bibnamefont
  {Niazi}}, \bibinfo {author} {\bibfnamefont {E.~V.}\ \bibnamefont
  {Sampathkumaran}}, \bibinfo {author} {\bibfnamefont {P.~L.}\ \bibnamefont
  {Paulose}}, \bibinfo {author} {\bibfnamefont {D.}~\bibnamefont {Eckert}},
  \bibinfo {author} {\bibfnamefont {A.}~\bibnamefont {Handstein}}, \ and\
  \bibinfo {author} {\bibfnamefont {K.~H.}\ \bibnamefont {M\"{u}ller}},\ }\href
  {\doibase 10.1103/PhysRevB.65.064418} {\bibfield  {journal} {\bibinfo
  {journal} {Phys. Rev. B}\ }\textbf {\bibinfo {volume} {65}},\ \bibinfo
  {pages} {064418} (\bibinfo {year} {2002})}\BibitemShut {NoStop}%
\bibitem [{\citenamefont {McClarty}\ \emph {et~al.}(2014)\citenamefont
  {McClarty}, \citenamefont {Hillier}, \citenamefont {Adroja}, \citenamefont
  {Kockelmann}, \citenamefont {Khalyavin}, \citenamefont {Wu}, \citenamefont
  {Manuel}, \citenamefont {Rayaprol},\ and\ \citenamefont
  {Sampathkumaran}}]{McClarty2014asm}%
  \BibitemOpen
  \bibfield  {author} {\bibinfo {author} {\bibfnamefont {P.}~\bibnamefont
  {McClarty}}, \bibinfo {author} {\bibfnamefont {A.}~\bibnamefont {Hillier}},
  \bibinfo {author} {\bibfnamefont {D.}~\bibnamefont {Adroja}}, \bibinfo
  {author} {\bibfnamefont {W.}~\bibnamefont {Kockelmann}}, \bibinfo {author}
  {\bibfnamefont {D.~D.}\ \bibnamefont {Khalyavin}}, \bibinfo {author}
  {\bibfnamefont {W.}~\bibnamefont {Wu}}, \bibinfo {author} {\bibfnamefont
  {P.}~\bibnamefont {Manuel}}, \bibinfo {author} {\bibfnamefont
  {S.}~\bibnamefont {Rayaprol}}, \ and\ \bibinfo {author} {\bibfnamefont
  {E.}~\bibnamefont {Sampathkumaran}},\ }\href@noop {} {\bibfield  {journal} {\bibinfo
  {journal} {unpublished}}}{\  (\bibinfo {year}
  {2014})}\BibitemShut {NoStop}%
\bibitem [{\citenamefont {Adroja}(2014)}]{Adroja2014sm}%
  \BibitemOpen
  \bibfield  {author} {\bibinfo {author} {\bibfnamefont {D.~T.}\ \bibnamefont
  {Adroja}},\ }\href@noop {} {\bibfield  {journal} {\bibinfo
  {journal} {unpublished}}}{\  (\bibinfo {year} {2014})}\BibitemShut
  {NoStop}%
\bibitem [{\citenamefont {Ashcroft}\ and\ \citenamefont
  {Mermin}(1976)}]{Ashcroft1976sm}%
  \BibitemOpen
  \bibfield  {author} {\bibinfo {author} {\bibfnamefont {N.~W.}\ \bibnamefont
  {Ashcroft}}\ and\ \bibinfo {author} {\bibfnamefont {N.~D.}\ \bibnamefont
  {Mermin}},\ }\href@noop {} {\emph {\bibinfo {title} {{Solid state
  Physics}}}}\ (\bibinfo  {publisher} {W. B. Saunders Company},\ \bibinfo
  {address} {Philadelphia},\ \bibinfo {year} {1976})\ p.\ \bibinfo {pages}
  {427}\BibitemShut {NoStop}%
\bibitem [{\citenamefont {Holstein}\ and\ \citenamefont
  {Primakoff}(1940)}]{Holstein1940sm}%
  \BibitemOpen
  \bibfield  {author} {\bibinfo {author} {\bibfnamefont {T.}~\bibnamefont
  {Holstein}}\ and\ \bibinfo {author} {\bibfnamefont {H.}~\bibnamefont
  {Primakoff}},\ }\href {\doibase 10.1103/PhysRev.58.1098} {\bibfield
  {journal} {\bibinfo  {journal} {Phys. Rev.}\ }\textbf {\bibinfo {volume}
  {58}},\ \bibinfo {pages} {1098} (\bibinfo {year} {1940})}\BibitemShut
  {NoStop}%
\bibitem [{\citenamefont {Bogoliubov}(1947)}]{Bogoliubov1947sm}%
  \BibitemOpen
  \bibfield  {author} {\bibinfo {author} {\bibfnamefont {N.}~\bibnamefont
  {Bogoliubov}},\ }\href@noop {} {\bibfield  {journal} {\bibinfo  {journal} {J.
  Phys.}\ }\textbf {\bibinfo {volume} {11}},\ \bibinfo {pages} {23} (\bibinfo
  {year} {1947})}\BibitemShut {NoStop}%

\end{thebibliography}
\end{document}